\documentclass[10pt,a4paper,twocolumn]{vupreprint}
\usepackage[bindingoffset=0.5cm,textheight=25cm,hdivide={2.5cm,*,3cm}, vdivide={*,24cm,*}]{geometry}
\usepackage{amsmath}
\usepackage{amsfonts}
\usepackage{amssymb}
 \usepackage[singlelinecheck=true,font=small]{caption}
 \usepackage[numbers,square,comma,sort&compress]{natbib}
 \usepackage{fancyhdr}
 \usepackage{fancybox}
 \usepackage{authblk}
 \usepackage{booktabs}
 \usepackage{multirow}
 \usepackage{soul}
 \usepackage{microtype}
 \usepackage{float}
 \usepackage[pdftex]{graphicx}
 \usepackage[pdftex,bookmarksnumbered=true,breaklinks=true]{hyperref}
\graphicspath{{}{./figures}}

\usepackage{upgreek}
\usepackage{textcomp}

\usepackage[cal=cm]{mathalfa}
\usepackage{braket}
\usepackage{sidecap}
\usepackage{color}
\usepackage{bm}
\usepackage{bbm}
\usepackage{tikz}
\usepackage{nicefrac}
\usepackage[utf8]{inputenc}
\usepackage[normalem]{ulem}
\usepackage{nicefrac}
\usepackage{sidecap}
\usepackage{xcolor, soul} 
\usepackage{physics}
\usepackage{mathrsfs,amsfonts,dsfont} 
\usepackage{braket}
\usepackage{enumerate}
\usepackage{ulem}
\usepackage{caption}
\usepackage[version-1-compatibility]{siunitx}
\usetikzlibrary{decorations.pathmorphing,decorations.pathreplacing,decorations.shapes}

\newcommand*{\doi}[1]{\href{https://doi.org/\detokenize{#1}}{doi: \detokenize{#1}}}
\newcommand*{\eprint}[1]{\href{https://arxiv.org/abs/#1}{arXiv:{#1}}}
\newcommand{\cyan}{\textcolor{black}} %
\newcommand{\blue}{\textcolor{black}} %
\newcommand{\red}{\textcolor{black}} %
\newcommand{\cut}[1]{}

\newcommand{\ri}{{\rm i}}						
\newcommand{\re}{{\rm e}}	

\newcommand{\eqnref}[1]{Eq.~(\ref{#1})}		
\newcommand{\figref}[1]{Fig.~\ref{#1}}			
\newcommand{\tabref}[1]{Tab.~\ref{#1}}		
\newcommand{\secref}[1]{Section~\ref{#1}}		

\newcommand{\qbounce}{{\it{q}}{\sc{Bounce}}}		
\newcommand{\inv}[1]{\frac{1}{#1}}					

\renewcommand{\a}{\alpha}

\renewcommand{\l}{\lambda}

\renewcommand{\Xi}{\Xi}

\definecolor{tublue}{cmyk}{1,0.38,0,0.15}
\newlength{\boxtextsept}
\newlength{\innertextwidth}
 \printdoi{10.1038/s42254-021-00298-2}
 \pacs{neutron physics, neutron interferometry, quantum foundations \& statistics, axions, non-Newtonian gravity}
 \journalref{S. Sponar, R. I.P. Sedmik, M. Pitschmann, H. Abele, and Y. Hasegawa}{Nat. Rev. Phys.}{3}{309 -- 327}{2021}{Tests of fundamental quantum mechanics and dark interactions\\with low-energy neutrons}
 \author{\red{Stephan} Sponar$^{1}$\footnote{stephan.sponar@tuwien.ac.at\\www.neutroninterferometry.com}}
\author{\cyan{Ren\'e} I.P. Sedmik$^1$}
\author{\blue{Mario} Pitschmann$^{1}$}
\author{Hartmut Abele$^1$}
\author{Yuji Hasegawa$^{1,2}$}
 \affil{$^1$Atominstitut, TU Wien, Stadionallee 2, 1020 Vienna, Austria}
\affil{$^2$Department of Applied Physics, Hokkaido University, Kita-ku, Sapporo 060-8628, Japan}
 
 \renewcommand{\preprt}{\textsc{Extended} {arXiv} \textsc{Version}}
\hypersetup{pdfauthor={\theauthors},pdftitle={\thetitle}}
\title{Tests of Fundamental Quantum Mechanics and Dark Interactions \\with Low Energy Neutrons - Extended Version}
\begin{document}
\twocolumn[
  \begin{@twocolumnfalse}
   \maketitle
\begin{abstract}
\begin{center}
\end{center}
Among the known particles, the neutron takes a special position, as it provides experimental access to all four fundamental forces and a wide range of hypothetical interactions. Despite being unstable, free neutrons live long enough to be used as test particles in interferometric, spectroscopic, and scattering experiments probing low-energy scales. 
As was already recognized in the 1970s,	fundamental concepts of quantum mechanics can be tested in neutron interferometry using silicon perfect-single-crystals.
Besides allowing for tests of uncertainty relations, Bell inequalities and alike, neutrons offer the opportunity to observe the effects of gravity and hypothetical dark forces acting on extended matter wave functions. 
Such tests gain importance in the light of recent discoveries of inconsistencies in our understanding of cosmology as well as the incompatibility between quantum mechanics and general relativity. Experiments with low-energy neutrons are thus indispensable tools for probing fundamental physics and represent a complementary approach to colliders. 
In this review we discuss the history and experimental methods used at this low-energy frontier of physics and collect bounds and limits on quantum mechanical relations and dark energy interactions.
\end{abstract}
\end{@twocolumnfalse}
]

\section{Introduction}
\label{sec:intro}

Wave-particle duality \cite{DeBroglie}, a particular form of complementarity \cite{BOHR1928}, addresses the incompatibility between the wave-like behavior of (massive) quantum objects and the `classical' concept of a particle. 
Depending on the experimental context, either wave or particle features can be probed.
In addition, as a massive, electrically neutral object with non-zero magnetic moment, the neutron gives experimental access to all four fundamental forces. That is, the strong interaction (due to its composition of quarks and gluons as well as in interactions with other nuclei), the weak interaction (as $\beta$-decaying particle), the electromagnetic force (mostly via the magnetic dipole moment), and gravitation. Neutrons are usually classified according to their energy range, from fast to slow (small to large wave length)~\cite{BookSears} as: \, i) relativistic, ii) fast, iii) epi-thermal, iv) thermal, v) cold, and vi) ultracold (see Box 1 for typical values of energy, wavelength and velocity and \eqnref{eq:schroedinger_generic} for non-relativistic Schr\"odinger equation). With its dual nature --- in some respects a wave, in other respects a particle --- the neutron is an almost ideal tool for studying both sides. 

{\red{}For a century, \emph{quantum mechanics} (QM) and its peculiarities arising from the wave-particle duality \cite{DeBroglie} as a particular form of complementarity \cite{BOHR1928}, and statistical interpretations have not only fascinated and puzzled generations of scientists, but have also become an indispensable topic in popular science. Prominent examples are Schr\"odinger's cat \cite{Schroedinger35}, Heisenberg's gamma-ray microscope \cite{Heisenberg27}, and the Einstein-Podolsky-Rosen (EPR) paradox \cite{Einstein35}.} The predictions of QM have been tested over and over again at the highest level of accuracy. Therefore, It can therefore be claimed that quantum mechanics is one of the best-verified theories in physics. Nonetheless, QM only gives probabilistic predictions for individual events. In order to maintain `a realistic world-view' \cite{Einstein35}, Einstein assumed that a more complete and deterministic theory must underlie quantum mechanics. Together with Podolsky and Rosen, Einstein developed the famous EPR experiments to test the predictions of QM against \emph{realistic} theories \cite{SHIMONY1973}. Numerous classes of such theories, for instance local realistic \cite{Bell66}, or non-contextual hidden variables \cite{Mermin93}, have been ruled out and no indications have been found for any contradiction of QM \cite{BookBell}. Eventually, a century after its inception, QM has found many practical applications and formed the basis for new technology now referred to as quantum computation and information technology \cite{NielsenChuang}.

\begin{figure*}[t]
\begin{tikzpicture}
  \node (box) [draw=tublue,thick,rounded corners=3pt,inner sep=\boxtextsept,text width=\innertextwidth,text justified,anchor=north]
  {\mbox{}\vspace{12pt}\\{\small
 Neutrons are usually divided into regions, depending on their energy: 



\vspace{5mm}

 \begin{center}
 \includegraphics[scale=0.9]{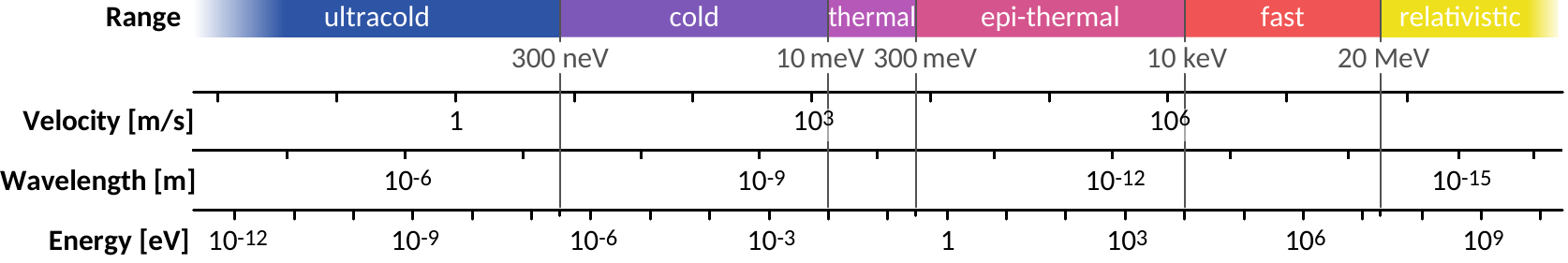}
 \captionsetup{labelformat=empty}
 \caption{ \label{fig:IFM}}
\end{center}

%
 At all energy scales relevant for this review, a neutron can be described by the non-relativistic Schr\"odinger equation 
 \begin{align}
 \left(-\frac{\hbar^2}{2m}\boldsymbol\nabla^2+V(\boldsymbol r,t)\right)\,\psi(\boldsymbol r,t)=i\hbar\frac{\partial}{\partial t}\psi(\boldsymbol r,t)\,.\label{eq:schroedinger_generic}
 \end{align}
Interactions with nuclei and matter are defined by the potential $V$. For nuclei $i$ located at $\boldsymbol r_{i}$, we have $V_{\rm{nuc}}(\boldsymbol r)={2\pi\hbar^2 b_{\mathrm{c}}}/{m_n}\sum_i\delta(\boldsymbol r-\boldsymbol r_{i,j})$ depending on the coherent scattering length $b_c$. At lower energies (cold and ultracold neutrons), the wavelength of the neutron exceeds the interatomic distances and only the optical (average pseudo-Fermi) potential $\bar V_{\rm{nuc}}=V_{\rm F}=2\pi\hbar^2 b_c \rho_N/m_n$ is important (see \secref{sec:grav:optics}). If magnetic fields are involved, we have in addition $V_{\rm{mag}}=-\boldsymbol{\mu}\cdot \boldsymbol{B}(\boldsymbol{r},t)=-\mu\; \boldsymbol{\sigma}\cdot \boldsymbol{B}(\boldsymbol{r},t)$ with the neutron's magnetic dipole moment $\boldsymbol \mu$. Gravity and hypothetical potentials have to be included, too, as described in \secref{sec:grav}.

\vspace{5mm}
Potentials and typical numbers for nuclear, gravitational and magnetic interactions:
\begin{center}
\begin{tabular}{l c l}
  \toprule
 \textbf{Interaction} & \textbf{Potential} & \textbf{Typical numbers} \\    
  \midrule
Nuclear (optical)  &  $2\pi\hbar^2 b_c\delta(\boldsymbol r)/m_n$ & $V_{\rm{Si}}\sim50$\,neV\\
Gravitational &  $m_n\,\boldsymbol{g}\cdot\boldsymbol{r}$  & $\sim$100\,neV per m\\
Magnetic & $-\boldsymbol{\mu}\cdot \boldsymbol{B}(\boldsymbol{r},t)$ & $\sim$60\,neV per T\\
  \bottomrule
\end{tabular}
\end{center}
with $\boldsymbol {B}:$ magnetic field strength, $\boldsymbol {g}:$ gravitational acceleration, $m_n$ neutron mass, $V_{\rm{Si}}$ optical (average pseudo-Fermi) potential of silicon.
}};
\node[draw=tublue,thick,inner sep=3pt,text width=\innertextwidth+14pt,text justified,anchor=north,fill=tublue!20] {\vspace{-11pt}
\mbox{}\hspace{3pt}\begin{minipage}{\innertextwidth}
\vspace{2pt}
 \subsubsection*{Box 1: The Neutron's Properties and Interaction with Matter and Fields }
\vspace{-6pt}\mbox{}
\end{minipage}
};
\end{tikzpicture}
\end{figure*}
Superposition of de Broglie matter waves, commonly referred to as \emph{matter-wave} interferometry, was first observed for electrons in the mid 1950ies~\cite{Elektron}, followed by thermal neutrons in 1970ies~\cite{Rauch74} and atoms a decade later~\cite{Atom}. Recently, even larger objects, like molecules (with masses exceeding 10000 amu) were used in interference experiments \cite{Eibenberger13}. 
Neutron interferometry \cite{Rauch74}, which makes use of the Bragg diffraction by successive slices cut from a single (perfect) crystal of silicon, provides a powerful tool for matter-wave interference. Neutrons are particularly well-suited for applications in matter-wave interferometry for several reasons: neutron interferometers offer a macroscopic beam separation (several centimeters), which allows manipulation of the individual sub-beams. Furthermore, neutrons can be detected with an efficiency close to one, and due to the relatively large de Broglie wavelength of thermal neutrons (see table in Box 1) neutrons can be manipulated coherently and efficiently over a long time ($\sim 10^3\,$s). Soon after its first experimental verification in 1974, neutron interferometry was used to demonstrate fundamental concepts of QM such as the spinor superposition law, quantum state entanglement, or 4$\pi$ symmetry of fermions. Several of these concepts were observed for the first time in the single neutron quantum system \cite{RauchBook}. More recent experiments demonstrated the entanglement between different degrees of freedom (spin and path)~\cite{Hasegawa03,Sponar10,Geppert14,Klepp14,Hasegawa10,Erdoesi13}, the contextual nature of quantum mechanics \cite{Hasegawa2006contextual,Bartosik09}, and the measurement of weak values~\cite{Denkmayr14,Sponar15,Denkmayr17,Geppert18}.

As every massive object, the neutron is affected by Newton's gravitational force. The resulting interaction not only appears in the trajectories of motion but also in the phase of the wave function, when the neutron is in a spatial superposition of different gravitational potentials. The first experimental demonstration of the associated gravitational phase shift was measured by Colella, Overhauser, and Werner (COW) in 1975~\cite{Colella75} using a neutron interferometer.  Soon afterwards the earth's rotation affecting the orbital angular momentum (Sagnac effect) \cite{Werner79}, and motional effect on the wave function (Fizeau effect) \cite{Arif89} were demonstrated.
In more recent experiments Newton's inverse square law of gravity was tested at micron distances on an energy scale of $10^{-14}$\,eV searching for hypothetical fifth forces. Gravity Resonance Spectroscopy (GRS) with ultracold neutrons ~\cite{Jenke:2011zz,Jenke:2014yel,Cronenberg:2018qxf}, neutron interferometry with thermal neutrons \cite{Lemmel:2015kwa}, and other experimental techniques described in \secref{sec:grav} provide constraints on many possible new interactions.

The largest part of the matter and energy distribution in our current universe is, as of now, completely unknown. In fact, the known particle content of the Standard Model amounts to only about 5 \% of the total mass. 
Dark matter (DM) has been postulated to account for roughly one third of the missing matter. Almost a century after the first experimental indications, the existence of DM as gravitationally interacting pressure-less `fluid' is well established~\cite{Bertone:2013} but its constitution is still unknown. Numerous hypothetical particles provide possible candidates for DM. A prime example is the axion~\cite{Weinberg:1977ma, Wilczek:1977pj, Kim:1979if, Shifman:1979if, Dine:1981rt}. This pseudoscalar particle was introduced as a consequence of an elegant mechanism that Peccei and Quinn proposed to solve the so-called strong CP problem~\cite{Peccei:1977hh, Peccei:1977ur}. Experimental constraints coming from astrophysics imply that the axion mass has to be very small ($\lesssim10$ meV \cite{Raffelt:1999tx}). This necessarily leads to long-range forces, which are, in principle, observable in laboratory experiments \cite{Moody:1984ba}. Astrophysical observations as well as terrestrial experiments exclude the possibility of heavier axions. Lighter axions, however, are are not yet experimentally excluded and are theoretically well motivated \cite{Freivogel:2008qc, Linde:1987bx}.
Although devised in a different context, axions provide a well-motivated DM candidate \cite{Bertone:2013, Raffelt:1999tx, Duffy:2009ig, Graham:2015ouw}.
A great number of other related proposals for light bosons are termed axion-like particles (ALPs), some of which are discussed in \secref{sec:grav:MGDMDE}.

While DM concerns the (missing) matter content associated with galaxies, the recently observed accelerated expansion of our universe lead to the postulation of a new substance of unknown origin termed `dark energy' (DE)~\cite{Perlmutter:2012, Riess:2012, Schmidt:2012xxa}. In GR, the standard theory describing our universe at cosmological scales, accelerated expansion can only be described when a so-called cosmological constant is included \cite{Einstein:1917}. Unfortunately, such an extension of Einstein's theory leads to a severe fine-tuning problem \cite{Sola:2013gha}. As is well-known, modifications of GR at cosmological scales are severely restricted, since they typically give rise to theoretical inconsistencies. Consequently, it appears more natural to add hypothetical new fields instead. A prominent proposal for such a scalar field is `Quintessence', which effectively provides a time-dependent cosmological `constant' via its potential energy~\cite{Joyce:2014kja}. Again, such fields would induce new interactions, so-called fifth forces, that are strongly constrained by many observational tests at solar distance scales and below \cite{Will:2014kxa}. To avoid those observational constraints in regions of comparably high mass density necessitates some kind of `screening mechanism', which hides the scalar interaction in regions where it would conflict with experimental bounds. Prominent among several different possible mechanisms are the so-called chameleon models \cite{Khoury:2003rn,Khoury:2003aq,Mota:2006ed,Mota:2006fz,Waterhouse:2006wv}. Extensive searches employing many experiments have put stringent bounds on the parameters of these models. A close relative of the chameleon is the symmetron, the effective potential of which resembles the Higgs potential of the Standard Model \cite{Hinterbichler:2010es, Hinterbichler:2011ca, Pietroni:2005pv, Olive:2007aj}. 

Both DM as well as DE models can be tested and searched for in laboratory experiments which are sensitive to the respective interaction potentials (see box 3). In fact, neutrons are perfectly suited to probing such potentials~\cite{Abele:2008zz,Brax:2011hb} and, consequently, provide an entrance into the dark sector.

The topic fundamental quantum mechanics and dark matter interactions with low energetic neutrons has been covered for instance by books, reviews and conference proceedings. Literature in some cases is too extensive to be cited and we therefore refer to recent reviews or papers and the references therein. Rauch and Werner discuss in their book “\emph{Neutron Interferometry, Lessons in Experimental Quantum Mechanics“}, wave-particle duality, coherence and decoherence, gravitationally induced quantum phase shifts, Berry’s geometrical phases, Aharonov-Bohm topological interference effects, and entanglement \cite{RauchBook}. They have put together a comprehensive compendium on quantum interference experiments. More recent fundamental phenomena of quantum mechanics explored with neutron interferometers, covering topics like quantum contextuality and multipartite entanglement, are reviewed by Hasegawa et al. in \cite{Klepp14}. Screended DE models are covered by a living review of Burrage and Sakstein \cite{Burrage:2017qrf}. They challenge `the gravitational inverse-square law' reviewed by Adelberger et al.~\cite{Adelberger:2009zz}, and Adelberger, Heckel and Nelson~\cite{Adelberger:2003zx}, and in the book by Fischbach~\cite{Fischbach:1999}. The review by Dubbers~\cite{Dubbers:2011ns} covers the neutron and its role in cosmology and particle physics, while the neutron as a particle with its properties and basic interactions was reviewed  by Abele~\cite{Abele:2008zz}. Conference proceedings of particle physics with neutrons, which cover topics of this review were edited by Jenke et al.~\cite{Jenke:2019proc}, Soldner et al.~\cite{Soldner:2008proc}, and earlier by Arif et al.~\cite{Arif:2005proc}, Zimmer et al.~\cite{Zimmer:2000proc}, Dubbers et al. ~\cite{Dubbers:1989proc}, Desplanques et al.~\cite{Desplanques:1984proc}, and von Egidy~\cite{Egidy:1978proc}. 

Our review is divided in two main parts. The first one describes tests of QM with an emphasis on neutron-based tests. In the second part, we give an overview of DM and DE models that have been tested using neutrons, relevant experimental methods, and the resulting limits on parameters.

\section{Tests of Quantum Mechanics}
\label{sec:qm_tests}

\begin{figure*}[t]
\begin{tikzpicture}
  \node[draw=tublue,thick,rounded corners=3pt,inner sep=\boxtextsept,text width=\innertextwidth,text justified,anchor=north]
  {\mbox{}\vspace{12pt}\\{\small
As already mentioned, neutrons are affected by all four fundamental interactions, which are utilized in realizing neutron optical components as shown in the \emph{tool-boxes} in \figref{fig:IFM} (b):  (i) phase shifters: by inserting a phase-shifter plate (or phase flag) into the interferometer, the phase relation between the two sub-beams, belonging to path $\mathrm{I}$ and $\mathrm{II}$ can be varied. The phase shift is given by $\chi=N_{\rm ps}b_{\rm c}\lambda D$ with atom density $N_{\rm ps}$ in the phase shifter plate of thickness $D$, the coherent scattering length $b_{\rm c}$ and the neutron wavelength $\lambda$. By rotating the phase shifter plate by an angle $\eta$, the phase shift $\chi$ can be tuned systematically, due to the change of the relative optical path length in path $\mathrm{I}$ and path $\mathrm{II}$. (ii) rotation stages: after rotation of the entire interferometer by an angle $\alpha$ around the incident beam direction by employing a tilting stage the neutron's split wave packets travel at different heights, and therefore at different gravitational potentials. 
This yields a dispersive phase shift which depends on the angle $\alpha$, the area $A$ enclosing the sub-beams and the wavelength $\lambda$. (iii) spin rotators: the neutron couples to magnetic fields via its permanent magnetic dipole moment $\boldsymbol \mu$, where static as well as time-dependent magnetic fields are used to realize arbitrary spinor rotations in due to the Larmor precession of the neutron's polarization vector. With time-dependent magnetic fields not only the neutron's spin but also its total energy can be manipulated due to photon absorption or emission. 
\setcounter{figure}{0}
 \begin{center}
 \includegraphics[scale=0.6]{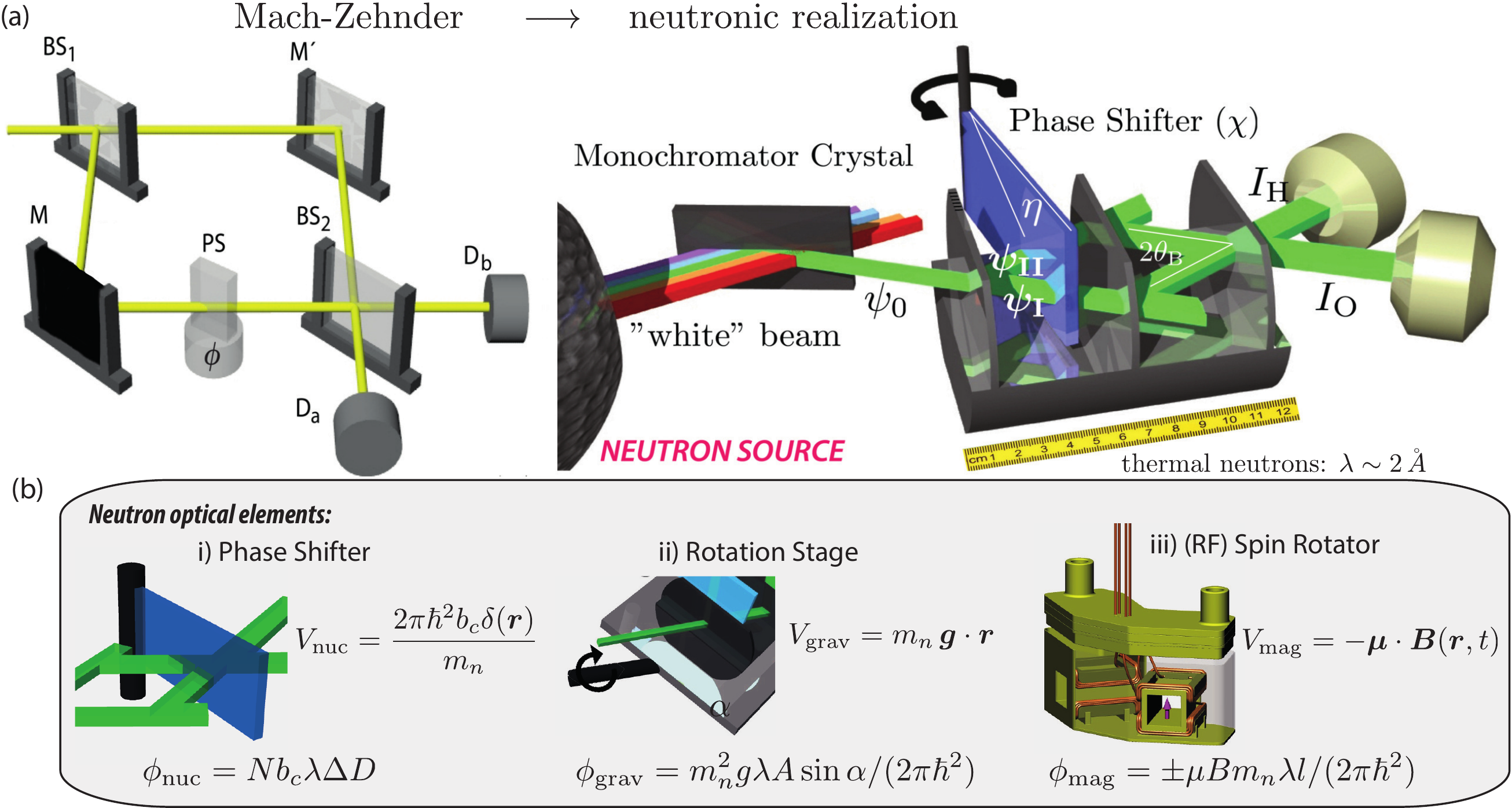}
 \caption{(a) Optical Mach-Zehnder interferometer compared to a silicon perfect crystal neutron interferometer of Mach-Zehnder type (triple Laue - LLL). (b) Tool box nuclear, gravitational and magnetic phase shifts.\label{fig:IFM}}
\end{center}
}};
\node[draw=tublue,thick,inner sep=3pt,text width=\innertextwidth+14pt,text justified,anchor=north,fill=tublue!20] {\vspace{-11pt}
\mbox{}\hspace{3pt}\begin{minipage}{\innertextwidth}
\vspace{2pt}
 \subsubsection*{Box 2. Single Perfect Crystal Neutron Interferometry}
\vspace{-6pt}\mbox{}
\end{minipage}
};
\end{tikzpicture}
\end{figure*}

\subsection{Neutron Interferometry and its Tools to Study Quantum Mechanics}
\label{sec:neutron_if}

In the 1960ies progress in semiconductor technology made it possible to produce monolithic perfect crystal silicon ingots, with a diameter of several inches, from which a single-crystal interferometer (based on knowledge for X-ray interferometry) can be cut out. In 1974 such an interferometer was illuminated with thermal neutrons, which resulted in the first observed interference fringes by Rauch, Treimer and Bonse \cite{Rauch74} at the rather small ($250\,$kW) TRIGA research reactor at the Atominstitut - TU Wien, in Vienna, Austria. In their seminal experiment, a beam of neutrons is split by amplitude division, and superposed coherently after passing through different regions of space. During these \emph{macroscopic} spatial separation, the neutron's wavefunction can be modified in phase and amplitude in various ways. In the interferometer, neutrons exhibit \emph{self-interference}, since at any given instant at most a single neutron propagates through the interferometer. The interferometer is geometrically analogous to the \textcolor{black}{well-known} Mach-Zehnder interferometer in photonics, as illustrated in \figref{fig:IFM}\,(a). It is worth noting here that a matter-wave interferometer (with completely separated sub-beams), was first realized by using neutrons; this is due to the following facts: (i) thermal neutrons have moderate de Broglie wavelength of $\sim 2\,\AA$ which is comparable to widely-used X-ray wavelength. (ii) charge neutrality of neutrons provides robustness against annoying-disturbance  from electronic interactions, which allows to maintain a higher degree of coherence for a prolonged time. 
This type of interferometer enables the realization of a quantum optical experiment employing matter-waves on a macroscopic scale; this has opened up a new era of investigations concerning fundamental quantum mechanical phenomena by employing matter-waves. In the following years numerous remarkable experiments on the foundations of QM have been carried out using neutron interferometry, e.g. the verification of the 4$\pi$ spinor symmetry \cite{Rauch754Pi}, which is nowadays applied in quantum gates where a $2\pi$ rotation of the applied system is performed to induce a $\pi$ phase shift. Furthermore, the superposition of different spin orientations of the two interferometer arms \cite{Badurek83Direct,Badurek83TimeDepend,Badurek86DoubleRes}, and the phase shift arising from a lateral confinement of a neutron beam passing through a narrow slit system \cite{Rauch:2002} have been investigated.

It is an ineluctable feature of quantum mechanics that an observed system is disturbed by any measurement. This fundamental principle of quantum mechanics is reflected in the famous Heisenberg uncertainty principle, published in 1927 \cite{Heisenberg27}, and illustrated in the famous gamma-ray-thought experiment. The uncertainty principle prohibits determination of certain pairs of quantum mechanical observables with arbitrary precision. In quantum mechanics a measurement process of an observable is described by
the projection postulate \cite{Neuman32}, where the outcome of the measurement is an eigenvalue of the observable and the initial state collapses into the corresponding eigenstate.

A different approach of the quantum mechanical measurement process, based on a two-vector formulation \cite{Aharonov64}, claims that the value of an observable depends on both a {\emph{pre-}} and a {\emph{post-selected}} state vector, respectively. This new kind of value for a quantum variable, referred to as \emph{weak value}, was introduced by Aharonov, Albert, and Vaidman (AAV) in 1988 \cite{Aharonov88}. The weak value of an observable of a quantum system can be determined via a procedure (called a \emph{weak measurement}), where some information of the quantum state can be extracted, without projecting the state into eigenstates. This is achieved by  weakly  coupling  the system  with  a  probe  system  (measurement device). 
 
 A topic closely related to weak measurements is \emph{entanglement}, which can be used to rule out certain classes of \emph{realistic theories} as possible extensions of QM.
Hidden variable theories are based on objective (realistic) properties of physical systems and were brought into play by Einstein and his co-workers Podolsky and Rosen (EPR). EPR claimed, based on the assumption of local realism, that quantum mechanics is not a complete theory \cite{EPR35}. In 1951, Bohm reformulated the EPR argument for spin observables of two spatially separated \emph{entangled} particles to illuminate the essential features of the EPR paradox \cite{BOHM51}. 
Following this reformulation, Bell proved in his theorem that all hidden variable theories, which are based on the joint assumption of locality and realism, conflict with the predictions of quantum mechanics \cite{Bell64}.

\subsection{Entanglement \& Contextuality}
\label{sec:qm_tests:entanglement}

\begin{figure*}[!t]
\begin{center}
\scalebox{0.6}
{\includegraphics{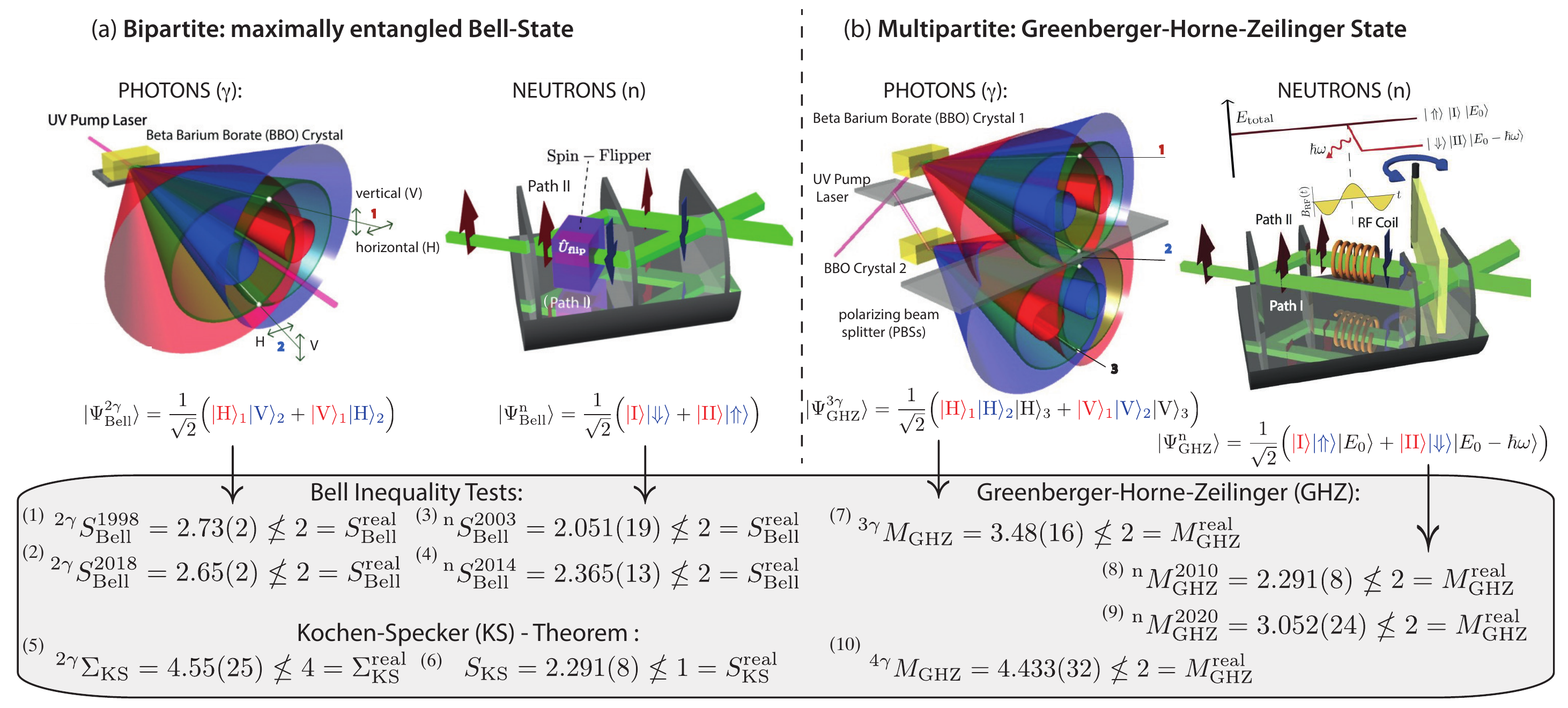}}
\caption{Entangled quantum state for photons ($\gamma$, multi-particle entanglement) in comparison with entangled single neutron quantum state (bipartite and tripartite intra-particle entanglement), used for (a) Bell tests (1) \cite{Weihs98}, (2) \cite{Rauch18}, (3) \cite{Hasegawa03} (4) \cite{Geppert14} and demonstration of the KS theorem (5) \cite{Cabello13} and (6) \cite{Bartosik09} (b) to probe GHZ entanglement (7) \cite{Pan2000}, (8) \cite{Hasegawa10}, (9) \cite{Shen2020} and (10) \cite{Erhard2018}. \label{fig:Entagnlement}}
\end{center}
\end{figure*}

The counter-intuitive nature of quantum mechanics, especially its probabilistic character, has been called into question from its very beginning. Einstein, Podolsky and Rosen (EPR) argued that  quantum mechanics is \emph{incomplete} with a hidden but more complete and deterministic physics underlying it \cite{EPR35}. In 1964 Bell proved in his celebrated theorem that all hidden-variable theories, or \emph{realistic} theories based on the assumptions of locality and realism lead to conflicts with the predictions of quantum mechanics \cite{Bell64}. Bell introduced inequalities that hold for the predictions of any local hidden-variable theory but are violated by quantum mechanics.

In parallel to Bell's work, Kochen and Specker (KS) devised another powerful argument, which shows that logical contradictions occur if non-contextual (\emph{realistic}) assumptions are considered \cite{Kochen67}. This theorem shows that the predictions of quantum mechanics are incompatible with the following assumptions: (i) a definite value of the measurements, i.e. observables $A$ and $B$ have predefined values $v(A)$ and $v(B)$ prior to a measurement and (ii) non-contextuality, i.e. the outcome of a measurement is assumed to be independent on the experimental context. This means that the measured value is the same, irrespective of whether any commuting observables are measured jointly. Quantum contextuality tests with neutrons are achieved by entangling the neutron's different degrees of freedom (DOF), which is referred to as \emph{intra-particle} entanglement and is depicted in \figref{fig:Entagnlement}\,(a). The first experimental realization, of a spin-path entangled neutron state was demonstrated by violating a Bell-like inequality reported in \cite{Hasegawa03}. The Bell-like single neutron state consists of a spin-path entanglement, which is realized by placing a spin flipper (depicted in \figref{fig:IFM}\,tool box) in one arm of the interferometer.  The neutron's state is  described  by  a  tensor  product  Hilbert  space $\mathcal H=\mathcal H_{\rm s}\otimes\mathcal H_{\rm p}$, where the former corresponds to the  spin wavefunction and the latter to the spatial wavefunction. Since observables of the spatial part commute with those of the spin part one can prepare a Bell-like state $\vert\Psi^{\rm n}_{\mathrm{\rm{Bell}}}\rangle=\frac{1}{\sqrt 2}\big(\vert\Uparrow\rangle \otimes\vert \mathrm{I}\rangle+\vert\Downarrow\rangle\otimes \vert \mathrm{II}\rangle\big)$, where $\vert\Uparrow\rangle $ and $\vert\Downarrow\rangle $ correspond to  spin up and the down eigenstates and $\vert \mathrm{I}\rangle$ and $\vert \mathrm{II}\rangle$ represent the two path eigenstates in the interferometer. Measurements of the joint expectation value of spin and path $E(\alpha,\chi)$, where $\alpha$ and $\chi$ denote the measurement directions of spin and path, respectively, yield a violation of the \emph{realistic} Bell-like inequality $-2 \leq S^{\rm{real}}_{\rm{Bell}}(\alpha,\chi) \leq2$. This is in contrast to the predictions of quantum mechanics, which yield $S^{\rm{QM}}_{\rm{Bell}}(\alpha,\chi) =2\sqrt 2$. In the actual experiment the figure of merit
\begin{eqnarray}
S_{\rm{Bell}}(\alpha,\chi)&=&E(\alpha_1,\chi_1+E(\alpha_1,\chi_2))\nonumber\\
&-&E(\alpha_2,\chi_1)+E(\alpha_2,\chi_2)\nonumber
\end{eqnarray}
is determined by applying the respective measurement direction for spin and path $\alpha_i$ and $\chi_i$ (with $i=1,2$). The results of the Bell experiments reported in  \cite{Hasegawa03} and  \cite{Geppert14}, where newly developed spin-rotators made it possible to obtain data with higher accuracy, are given in the respective gray box in \figref{fig:Entagnlement} (b).

Furthermore, in \cite{Hasegawa06} a test of the KS theorem was carried out with a neutron interferometer, where a combination of six observables has been evaluated \cite{Cabello08,Bartosik09}.  A contradiction with the predictions of  non-contextual hidden variable theories was obtained due to the contextual nature of quantum mechanics. For the proof of the KS theorem, we once again consider single neutrons prepared in a maximally entangled Bell-like state denoted as $\vert^-\Psi^{\rm n}_{\mathrm{\rm{Bell}}}\rangle=\frac{1}{\sqrt 2}\big(\vert\Uparrow\rangle \otimes\vert \mathrm{I}\rangle-\vert\Downarrow\rangle\otimes \vert \mathrm{II}\rangle\big)$. The proof is based on the six observables $\hat\sigma_{x}^{s}$, $\hat\sigma_{x}^{p}$, $\hat\sigma_{y}^{s}$, $\hat\sigma_{y}^{p}$, $\hat\sigma_{x}^{s}\hat\sigma_{y}^{p}$ and $\hat\sigma_{y}^{s}\hat\sigma_{x}^{p}$. 
The inconsistency between a non-contextual hidden variable theory and quantum mechanics arises in any attempt to ascribe the predefined values $-1$ or $+1$ to each of the six observables. 
Since experiments due to their finite precision never show perfect (anti-) correlations, one has to derive an inequality, which can be experimentally tested. It can be shown that in any non-contextual hidden variable theory the relation 
\begin{equation}
S^{\rm{real}}_{\rm{KS}}=-\langle\hat \sigma ^{s} _{x} \cdot\hat \sigma ^{p} _{x} \rangle - \langle \hat\sigma ^{s} _{y} \cdot\hat \sigma ^{p} _{y} \rangle  - \langle \sigma^{s} _{x} \hat\sigma ^{p} _{y} \cdot \hat\sigma ^{s} _{y} \hat\sigma ^{p} _{x} \rangle \leq 1\nonumber
\end{equation}
 holds, in contrast to the quantum mechanical prediction $S^{\rm{QM}}_{\rm{KS}}=3$. A violation of this inequality directly reveals quantum contextuality.  The final experimental results in favor of the quantum mechanical predictions, thereby ruling out non-contextuality, from \cite{Bartosik09}  are given in the gray box in \figref{fig:Entagnlement} (b).

However, in addition to the spin and path DOF also the (total) energy of a neutron can also be entangled, generating tri-partite entangled states like the so-called  Greenberger - Horne - Zeilinger (GHZ) state \cite{GHZ89Pro,GHZ90}, which has already been generated in a neutron interferometer. \textcolor{black}{Using a radio-frequency spin-flipper in one path of the interferometer one can manipulate the neutrons' total energy, thereby realizing a triple entanglement between the path, spin, and energy DOF \cite{Hasegawa10} with the states denoted as  $$\vert\Psi^{\rm n}_{\mathrm{\rm{GHZ}}}\rangle=\frac{1}{\sqrt 2}\big(\vert\Uparrow\rangle \otimes\vert \mathrm{I}\rangle \otimes\vert E_0\rangle+\vert\Downarrow\rangle\otimes \vert \mathrm{II}\rangle \otimes\vert E_0-\hbar\omega\rangle\big).$$ The setup is schematically illustrated in \figref{fig:Entagnlement} (b). In the 1990ies,  Mermin  analyzed  the  GHZ  argument  in  detail  and derived  an  inequality  suitable  for  experimental  tests  to distinguish  between  predictions  of  quantum mechanics  and  realistic theories \cite{Mermin90}. According to non-contextual hidden variable theories, one can calculate a limit for the sum of certain observables' expectation values, which can be tested in an experiment. For the expectation value of 
\begin{equation}
M_{\rm GHZ}=\hat\sigma_x^{\rm s}\hat\sigma_x^{\rm p}\hat\sigma_x^{\rm e}-\hat\sigma_x^{\rm s}\hat\sigma_y^{\rm p}\hat\sigma_y^{\rm e}-\hat\sigma_y^{\rm s}\hat\sigma_x^{\rm p}\hat\sigma_y^{\rm e}-\hat\sigma_y^{\rm s}\hat\sigma_y^{\rm p}\hat\sigma_x^{\rm e}\nonumber
\end{equation}
non-contextual hidden variable theories set a maximum limit of $M^{\rm{real}}_{\rm GHZ}=2$. In contrast, quantum theory predicts an upper bound of $M^{\rm{QM}}_{\rm GHZ}=4$. Consequently, any measured value of $M_{\rm GHZ} > 2$ decides in favor of quantum contextuality. The results obtained in \cite{Hasegawa10} are given in the gray box of \figref{fig:Entagnlement}. (b)} and again display the inconsistency between the quantum mechanical predictions and a non-contextual (\emph{realistic}) model. 

In a more recent experiment, the \emph{Larmor} spin-echo instrument located at the second target station of the ISIS neutron and muon source, which is part of Rutherford Appleton Laboratory (RAL) in the UK, has been used to perform Bell tests and to study GHZ states \cite{Shen2020}. The main difference to the interferometric approach is that in a neutron spin-echo interferometer the trajectories of the individual spin states are manipulated by using refraction through magnetic fields. The thereby induced spatial separation of trajectories ranges from nanometers to microns, while the beam separation in a silicon-perfect crystal interferometer is of the order of several centimeters (see table in \figref{fig:Entagnlement} for results). 

Quantization of Orbital Angular Momentum (OAM) - another DOF - of bound massive particles and free photons was discovered a long time ago. However, in recent years stable OAM has also been observed for electrons in terms of so-called \emph{electron vortex beams} \cite{Verbeeck10}. Furthermore, it was demonstrated that \emph{mixed} OAM states can be prepared in free thermal neutrons using a spiral phase plate \cite{Clark15}  and magnetic gradients \cite{Sarenac19}. However, with these methods it is not possible to create \emph{pure} OAM states, due to the small coherence length of thermal neutrons. OAM of mixed type are usually referred to as \emph{extrinsic} OAM, since each neutron in the beam has the same OAM with respect to the optical beam axis but a different \emph{intrinsic} OAM. So a different experimental approach to generate pure OAM states is required. Among the theoretically developed new methods for generating OAM in neutrons are magnetic quadrupoles that can create neutron OAM entangled to the neutron spin (spin-orbit states) \cite{Nsofini16}, a series of perpendicular linear magnetic gradients forming a spin-orbit lattice, parity violating neutron-nucleus interactions where the neutron spin is rotated slightly around the momentum vector due to the weak interaction \cite{Heckel84} and static homogeneous electric fields since in an electric field the particle spin couples to the cross product between the electric field strength and the particle momentum. 
This procedure involves the preparation of a spin-orbit textured \emph{lattice of vortices} wavefront. A variety of textures can be generated using this method \cite{Sarenac19}, including skyrmion-like geometries. In \cite{Geerits20} we present an experimental procedure, where a static homogeneous electric field, polarized along the direction of particle propagation induces longitudinal spin-orbit states,  while a transversely polarized electric field  generates  transverse  spin-orbit  states.   The  latter type of OAM has not yet been observed in massive free particles. OAM is foreseen to be utilized in  multipartite entanglement, where \emph{four} neutron DOF, namely spin, path, total energy and OAM contribute to a multipartite entangled single neutron state.

\subsection{Weak Values \& Weak Measurements}
 \label{sec:qm_tests:weak}
\begin{figure*}[!t]
\begin{center}
\scalebox{0.6}
{\includegraphics{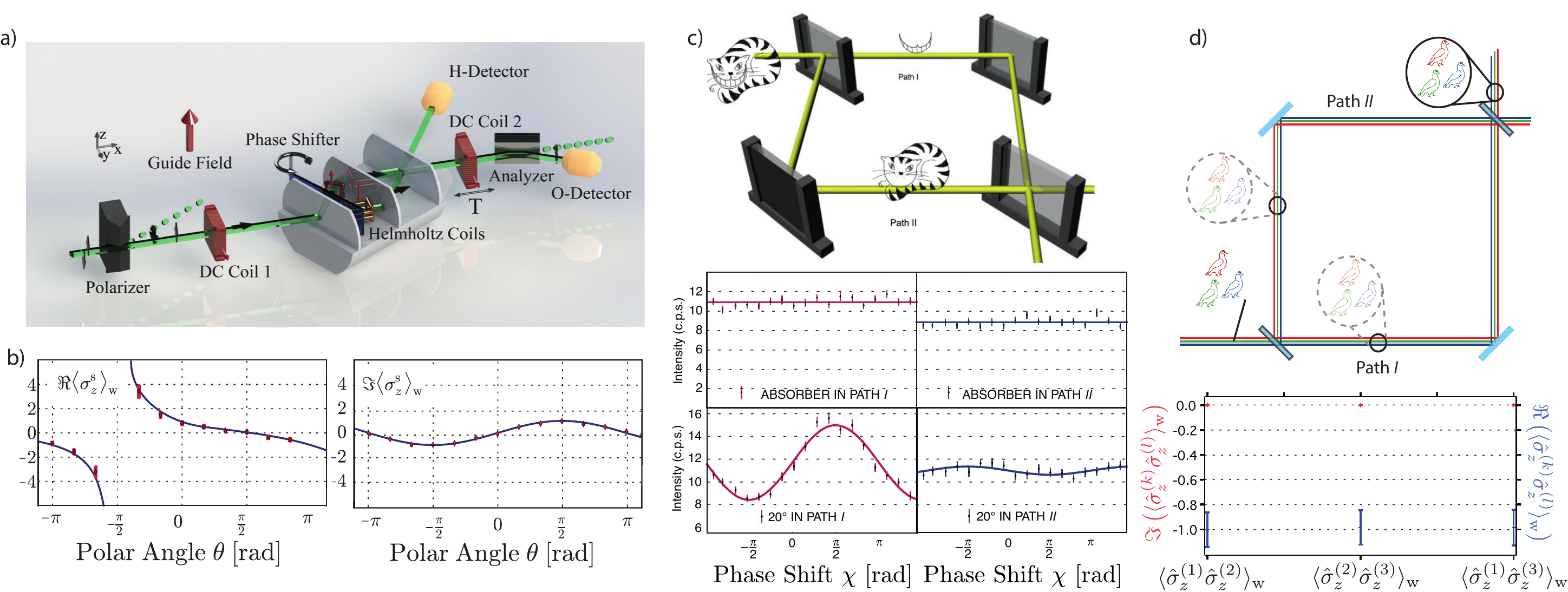}}
\caption{(a) Setup for measurement spin and path weak values (b) Real and imaginary component of weak measurement of Pauli spin operator $\sigma_z$ versus the polar angle $\theta$ and two selected azimuthal angles $\phi$ of the final (post-selected) spin states $\vert \psi_{\rm f}(\theta,\phi)\rangle=\cos(\theta/2) \vert\Uparrow\rangle+e^{{\rm i}\phi} \sin(\theta/2) \vert\Downarrow\rangle$ and initial (pre-selected) state $\vert \psi_{\rm i}\rangle=1/\sqrt 2(\vert\Uparrow\rangle+\vert\Downarrow\rangle)$. (c) Scheme and interferograms of quantum Cheshire Cat. (d) Scheme and results of quantum Pigeon Hole Principle.}\label{fig:weak}
\end{center}
\end{figure*}
\textcolor{black}{Another example of counter-intuitive prediction of quantum mechanics concerns the introduction of a new concept of quantum measurements, i.e., weak measurements with the resulting weak values,} as first proposed by Yakir Aharonov in 1988 \cite{Aharonov88}. The first corresponding experiment was realized using an optical setup \cite{Ritchie91}. The peculiarity of the weak value is that its value may lie far outside the range of an observable's eigenvalues and that it allows information to be extracted from a quantum system with only minimal disturbance. 
The weak value of a Hermitian operator $\hat A$ is defined as 
\begin{equation}
\langle \hat A\rangle _{\rm w}=\frac{\bra{\psi_{\rm f}}\hat A \ket{\psi_{\rm i}} }{\braket{\psi_{\rm f}}{\psi_{\rm i}}}  \nonumber
\end{equation}
and the corresponding  \emph{weak measurement} involves the three steps: (i) preparation of an initial quantum state $\vert\psi_{\rm i}\rangle$ (pre-selection) of the system; (ii) a weak coupling of this system with a probe, that is another quantum system and described by an interaction Hamiltonian. The interaction is supposed to be sufficiently weak, so that the system is only minimally disturbed; (iii) post-selection of the final quantum state $ \ket{\psi_{\rm f}}$, by performing a standard projective measurement of another observable $\hat B$ of the system. Finally, a measurement on the probe is performed, usually referred to as pointer read-out, yielding the weak value $\langle\hat A\rangle_{\rm w}$.

The weak value has been found to be useful as a technique which aims to amplify small signals \cite{Hosten08,Dixon09,Starling10,Starling10PRA,Feizpour11,Zhou13}, also demonstrated with neutrons (see  \figref{fig:weak}\,(a) for setup), where all aspects  of the weak value of the neutron's Pauli spin operator $\hat\sigma_z$, i.e.,  its real component, as well as the imaginary component, are experimentally determined, as reported in \cite{Sponar15}. The real part of the weak value exhibits values lying outside the usual range of spin eigenvalues, i.e.,$\pm 1$, ranging from -3.2 to 3.4. Furthermore, we observe non-zero values for the imaginary part of the weak value $\hat\sigma_z$, which can be seen in \figref{fig:weak}\,(b). Our results are an unambiguous quantum mechanical effect, since no classical theory can describe the observed weak measurement results. Our new experimental scheme allows full determination of the weak value, which can be used to characterize the evolution of the neutron's wave-function inside an interferometer, just like in a report of a photonic double-slit experiment \cite{Steinberg12}.

Another application of the weak value manifests concerns the estimation of quantum states \cite{Lundeen11,Goggin11,Steinberg12,Salvail13,Kaneda14,Ringbauer14,Denkmayr17}  (see \cite{Dressel14} for a recent review). Quantum tomography is a well-known approach to reconstruct a quantum state. However, quantum state tomography involves a lot of computational data postprocessing. In 2011 a novel more direct tomographical method was established that allows to determine a quantum state via a weak measurement without the postprocessing. However, that novel method had a drawback. Because of this weakness information gain is very low for each measurement and the measurements have to be repeated several times. We have now succeeded in combining both methods and as such obtain the benefits from both. We are able to employ the method established in 2011 without the need of computational postprocessing. At the same time we have managed to use strong measurements, thereby significantly reducing the  measurement time and enabling to determine the quantum state with higher precision and accuracy. We have performed a neutron interferometric experiment \cite{RauchBook,Sponar17}, the results obtained are not limited to that particular quantum system but are generally valid and as such may be applied to many other quantum systems as well.

In addition, weak values and weak measurements have been successfully applied to quantum paradoxes such as the three-box problem \cite{Steinberg04}, Hardy's paradox \cite{Steinberg09,Yokota09,Aharonov02PLA}, the pigeon principle \cite{Cai17} and the quantum Cheshire Cat \cite{Aharonov13,Denkmayr14}. The last two have been investigated with neutron interferometric experiments. Before presenting our experimental results we will introduce the quantum Cheshire Cat in more detail. \emph{"Well! I've often seen a cat without a grin," thought Alice; "but a grin without a cat! It's the most curious thing I ever saw in all my life!"}, these are Alice's famous words after finding a passage to a surreal world in a rabbit hole, where she meets a cat that leaves her wondering. Such a phenomenon, which at first seems absurd, is actually possible in a quantum mechanical sense for the quantum Cheshire Cat in a Mach-Zehnder interferometer. Here the cat itself is located in one beam path, while its grin is located in the other \cite{Aharonov13}. An artistic depiction of this behavior can be seen in \figref{fig:weak}\,(c). In our realization of the quantum Cheshire Cat, the neutron's path plays the role of the cat and the cat's grin is represented by the neutron's spin component along the $z$-direction. The system is initially prepared in an entangled state, given by $\vert\Psi_{\mathrm{i}}\rangle=\frac{1}{\sqrt 2}\big((\vert \Uparrow\rangle+\vert \Downarrow\rangle) \vert \mathrm{I}\rangle+(\vert \Uparrow\rangle-\vert \Downarrow\rangle)  \vert \mathrm{II}\rangle\big)$. For an observation of the quantum Cheshire Cat, after pre-selection of an ensemble, a weak measurement of the neutrons' population in a given path on the one hand and of the value of the spin in a given path on the other is performed. Subsequently, the ensemble is post-selected in the final state, which is the product state $\vert\Psi_{\mathrm{f}}\rangle=\frac{1}{\sqrt 2}(\vert\Uparrow\rangle+\vert\Downarrow\rangle)( \vert \mathrm{I}\rangle+ \vert \mathrm{II}\rangle)$. The weak values of the projection operators on the neutron path eigenstates $\hat\Pi_j=\ketbra{j}{j}$, with $j={\rm I}$ and ${\rm II}$ yield $\langle\hat\Pi_{\rm I}\rangle_{\rm w}=0$ and $\langle\hat\Pi_{\rm II}\rangle_{\rm w}=1$. The first expression indicates, that a weak interaction coupling of the spatial wavefunction to a probe localized on path $\mathrm{I}$, has no effect on the probe on average - the system behaves as if there were no neutron traveling on path $\mathrm{I}$. The weak value of the spin component along each path $j$ suggests the location of the neutrons' spin component. The appropriate observable of neutrons' spin component in path $j$ is given by $\langle\hat\sigma_z\hat\Pi_{j}\rangle_{\rm w}$, which yields $ \langle\hat\sigma_z\hat\Pi_{\rm I}\rangle_{\rm w}=1$ and $\langle\hat\sigma_z\hat\Pi_{\rm II}\rangle_{\rm w}=0$, for path $\mathrm{I}$ and $\mathrm{II}$, respectively. The experimentally obtained values of the weak measurements are determined by the observed intensities, plotted in \figref{fig:weak}\,(c), are explicitly given by $\langle\hat\Pi_{\rm I}\rangle_{\rm w}=0.14(4),\, \langle\hat\Pi_{\rm II}\rangle_{\rm w}=0.96(6),\,\vert\langle\hat\sigma_z\hat\Pi_{\rm I}\rangle_{\rm w}\vert^2=1.07(25)$, and  $\vert\langle\hat\sigma_z\hat\Pi_{\rm II}\rangle_{\rm w}\vert^2=0.02(24)$, for path and spin, respectively. Results have been published in \cite{Denkmayr14}. Later photonic realizations of the quantum Cheshire Cat are reported in \cite{Ashby16,Atherton15}.

The pigeonhole principle states that if $n$ pigeons are put into $m$ pigeonholes, with $n>m$ then at least one pigeonhole must contain more than one pigeon. However, in quantum mechanics this does not hold \cite{Aharonov16}. We consider 3 independent neutron spins and prepare them in the product state $|\psi\rangle = (|\Uparrow\rangle+|\Downarrow\rangle)^{\otimes 3}$ and post-select onto the product state  $|\phi\rangle = (|\Uparrow\rangle+\rm i|\Downarrow\rangle)^{\otimes 3}$. For our specific case of N = 3 spins, we consider a product of any two spin operators $\hat\sigma_z\hat\sigma_z$, with spectral decomposition $\hat\sigma_z\hat\sigma_z=(+1)\hat\Pi_{\rm{even}}+(-1)\hat\Pi_{\rm{odd}}$,  in terms of the rank-2 projectors, given by  $\hat\Pi_{\rm{even}}=\hat\Pi_+\otimes\hat\Pi_+ +\hat \Pi_-\otimes\hat\Pi_-$ and  $\hat\Pi_{\rm{odd}}=\hat\Pi_+\otimes\hat\Pi_- + \hat\Pi_-\otimes\hat\Pi_+$, with projection operators $\hat\Pi_\pm=\frac{{1\!\!1}\pm\hat\sigma_z}{2}$. Given pre- and post-selected states $|\psi\rangle$ and $|\phi\rangle$, as defined above, we get $(\hat\Pi_{\rm even})_{\rm w} = 0$ and $(\hat\Pi_{\rm odd})_{\rm w} = 1$, and thus $(\hat\sigma_z\hat\sigma_z)_{\rm w} = -1$, which implies $\hat\sigma_z\hat\sigma_z = -1$ for all pairs of spins according to the Aharonov-Bergmann-Lebowitz (ABL) formula \cite{ABL64}. This pairwise constraint is the quantum pigeonhole effect. To see this, let the spin eigenstates correspond to two boxes in which pigeons may be placed. $\hat\Pi_{\rm{even}}$ denotes two pigeons in one box (same spins) whereas $\hat\Pi_{\rm{odd}}$ denotes one pigeon in each (different spins). Since all three of our spins are pairwise anti-correlated no two pigeons are ever in the same box. The experimentally obtained values of the weak measurements of individual spins $(\hat\sigma_z)_{\rm w}$ and the pairwise anti-correlated spins $(\hat\sigma_z\hat\sigma_z)_{\rm w}$ are reported in \cite{Cai17} and plotted in Fig.\ref{fig:weak}\,(d). A photonic version of the quantum pigeonhole effect is reported in \cite{Chen19}.

As already discussed in Sec.\,\ref{sec:qm_tests:entanglement}, \emph{quantum contextuality}, as introduced by Kochen and Specker (KS) \cite{Kochen67}, forbids all observable properties of a system from being predefined independently of how they are observed. The KS theorem is proved by exhibiting a KS set of observables that contains geometrically related and mutually commuting subsets (or measurement contexts) that result in a logical incompatibility: Any non-contextual hidden variable theory (NCHVT) preassigning eigenvalues to the entire KS set, i.e., \emph{noncontextually}, results in a contradiction with the predictions of quantum mechanics. That is, at least one eigenvalue in a global assignment to a KS set cannot be predefined without violating a constraint on the product of eigenvalues within some context. A non-contextual hidden variable theory that assigns an eigenvalue prediction $\pm 1$ to each of the $3N$ observables must violate at least one of these product predictions. Crucially, such forbidden KS value assignments manifest as \emph{anomalous projector weak values}, meaning with real part outside the usual range $[0,1]$. On the other hand, classical assignments of pigeons to boxes must respect the range $[0,1].$  Consequently, anomaly indicates contradictory non-contextual value assignments to the corresponding context. Thus, the forbidden projectors thus constitute witness observables such that negative weak values imply confined KS contextuality and contradict the assignment of 0 by an NCHVT. In  \cite{Cai17} an unbiased \emph{contextuality observable} $C^{(N)}$ using all $2^{N-1}$ rank-2 projectors in an $N$-spin contextual basis is constructed as $C^{(N)} = I - \sum_{i=1}^{2^{N-1}} s_i\, \Pi^{(N)}_i $, with $s_i =  \textrm{sign}[\text{Re}(\Pi_i^{(N)})_w]$ and experimentally tested for all odd numbers of spins from $N=3$ to $N=17$. The corresponding \emph{pre-selected} product states is given by $|\psi_N\rangle = (|\Uparrow\rangle+|\Downarrow\rangle)^{\otimes N}$ and \emph{post-selected} product states by $|\phi_N\rangle = (|\Uparrow\rangle+{\rm{i}}|\Downarrow\rangle)^{\otimes N}$, respectively. Final results for all $N$ from 3 to 17 violate the non-contextuality bound $C^{(N)}_{\rm w} > 0$. The data for $N=5$ is most statistically significant, with $C^{(5)}_{\rm w} = -2.85 \pm 0.41$ violating the bound for NCHVTs by more than $7\sigma$. 

 \begin{figure*}[!t]
  \includegraphics[width=175mm]{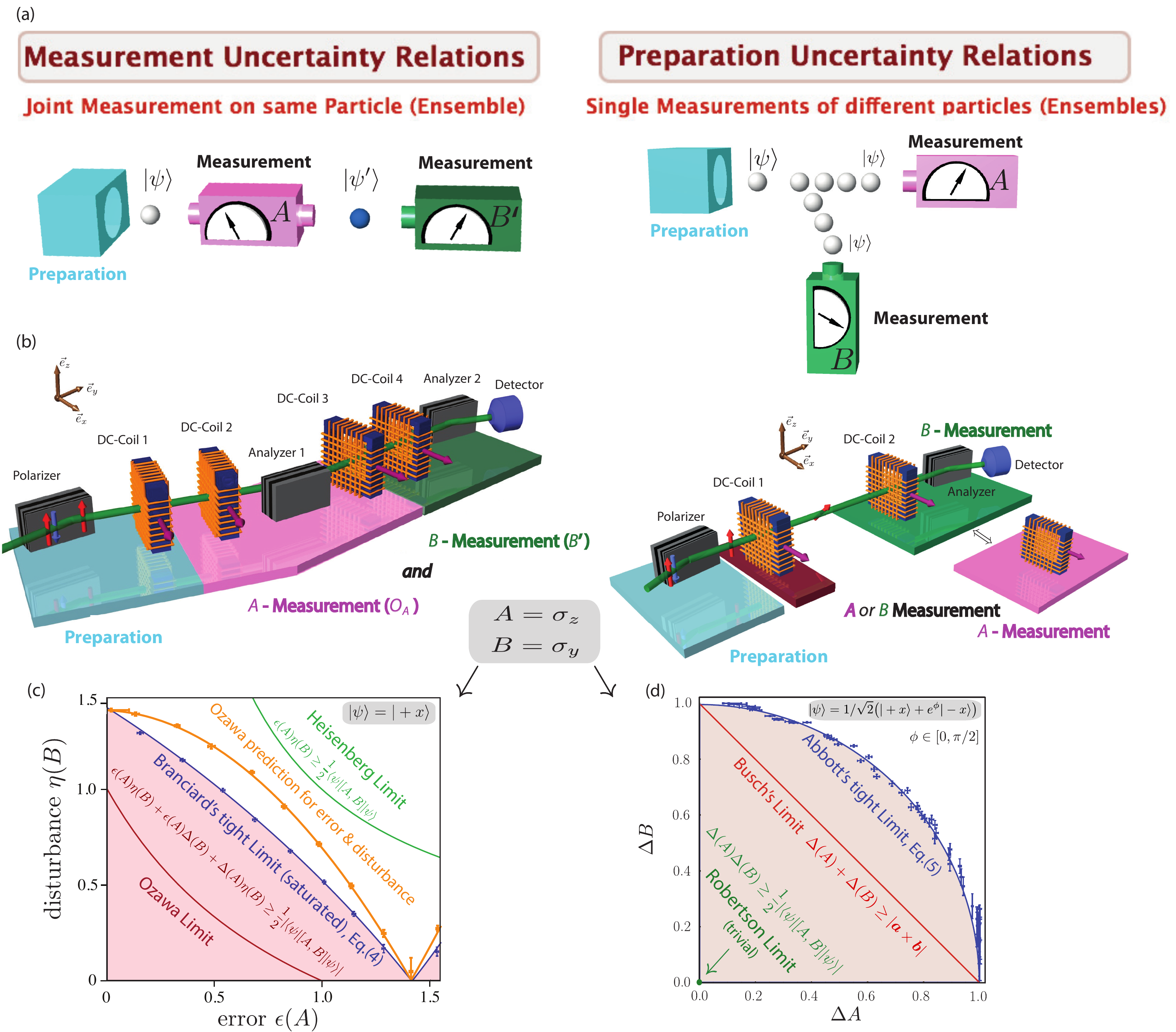}
  \caption{(a) Schematic illustration of measurement uncertainty (left) vs. preparation uncertainty measures. (b) Neutron optical setup for measurement uncertainty relation in successive spin measurements (left) and setup for preparation uncertainty relations (right). (c) Experimental results for Ozawa’s and Branciard’s tight measurement uncertainty relations, orange and blue data points, respectively, together with the Branciard bound from Eq. (4) (blue curve), bound from Ozawa’s relation Eq. (3) (red curve); green curve: bound imposed by Heisenberg's original error-disturbance relation  $\epsilon(A)\eta(B)\geq\frac{1}{2}\vert\bra{\psi} [A,B] \ket{\psi}\vert$ (panel d). (d) Experimental results of Abbott’s tight and state-independent preparation uncertainty from Eq. (5), together with Busch’s untight bound in red. Here, observables $A$ and $B$ are represented by Pauli spin matrices $\sigma_z$ and $\sigma_y$, $\Delta(X)$ denotes the standard deviation of an observable $X$, $\epsilon(A)$ and $\eta(B)$ account for error of $A$ and disturbance of $B$, respectively.  }\label{fig:Uncertainty}
\end{figure*}

\subsection{Uncertainty Relations}
\label{sec:qm_tests:uncertainty}

The uncertainty principle is without any doubt one of the cornerstones of modern quantum physics. In his original paper from 1927 \cite{Heisenberg27} Heisenberg proposed a reciprocal relation for the \emph{mean error} of a position measurement and \emph{disturbance} on the momentum measurement, thereby predicting a lower bound on the uncertainty of a joint measurement of incompatible observables. On the other hand this relation also sets an upper bound on the accuracy with which the values of non-commuting observables can be simultaneously prepared. While in the past these two statements have often been mixed, they are now clearly distinguished as \emph{measurement uncertainty} and \emph{preparation uncertainty} relations. Heisenberg was inspired by Einsteins realistic view to base a new physical theory only on observable quantities (\emph{elements of reality}), arguing that terms like velocity or position make no sense without defining an appropriate apparatus for a  measurement. By considering solely the Compton effect Heisenberg proposed the famous gamma-ray microscope thought-experiment. It gives a rather heuristic estimate for the product of the inaccuracy (error) of a position measurement $p_1$ and the disturbance $q_1$ induced on the particle\rq{}s momentum, denoted as $p_1q_1\sim h$. 
 \setcounter{equation}{1}

Heisenberg's paper presented his idea only heuristically. The first rigorously-proven  uncertainty relation for position $Q$ and momentum $P$ was introduced by Kennard \cite{Kennard27} as $\Delta(Q)\Delta(P)\geq\frac{\hbar}{2}$, in terms of \emph{standard deviations} defined as $\Delta^2(A)=\langle\psi\vert A^2\vert\psi\rangle-\langle\psi\vert A\vert\psi\rangle^2$ for an observable $A$. However, the $\gamma$-ray microscope sets a lower bound for the product of the measurement error and the disturbance in a joint measurement of position $Q$ and momentum $P$. Hence, the position-momentum uncertainty relation in terms of standard deviations quantifies how precise a state can be prepared with respect to the observables of interest. In 1929, Robertson \cite{Robertson29} extended Kennard's relation to arbitrary pairs of observables $ A$ and $ B$  as 
\begin{equation}\label{eq:Robertson}
\Delta( A)\Delta( B)\geq\frac{1}{2}\vert\langle\psi\vert[ A, B]\vert\psi\rangle\vert,
\end{equation}
with commutator $[ A, B]= A B- B A$. The corresponding generalized form of Heisenberg's original  error-disturbance uncertainty relation (measurement uncertainty relation) would read $\epsilon(A)\eta(B)\geq\frac{1}{2}\vert\bra{\psi} [A,B] \ket{\psi}\vert$. However, certain measurements do not obey this relation, as was pointed out first by Arthurs in 1988 \cite{Arthurs88}. Consequently, in 2003 Ozawa introduced the correct form of a generalized error-disturbance uncertainty based on rigorous theoretical treatments of quantum measurements \cite{Ozawa03,Ozawa04}
\begin{equation}\label{eq:Ozawa}
\epsilon(A)\eta(B)+ \epsilon(A)\Delta(B)+\Delta(A)\eta(B)\geq\frac{1}{2}\vert\bra{\psi} [A,B] \ket{\psi}\vert.
\end{equation}
In this relation, $\epsilon(A)$ denotes the root-mean-square (r.m.s.) error  of an arbitrary measurement for an observable $A$, $\eta(B)$ is the r.m.s. disturbance on another observable $B$ induced by the measurement, and $\Delta(A)$ and $\Delta (B)$ are the standard deviations of $A$ and $B$ in the state $\ket{\psi}$ before the measurement. The first measurement of observable $A$ (with error $\epsilon(A)$) disturbs the subsequent second measurement of observable $B$, modifying it to an effective $B'$ measurement (with disturbance $\eta(B)$), due to the change of the initial state from $\vert \psi\rangle$ to $\vert \psi'\rangle$, which is caused by the first measurement. In principle the first measurement can be done with arbitrary accuracy, that means $\epsilon(A)=0$, however in that case the disturbance of the second measurement will be maximal such that no information can be gained from the second measurement. Hence, instead of exactly measuring $A$ an approximate measurement, denoted as $O_A$ is performed, aiming to reduce the disturbance on the second measurement.   
Although universally valid, Ozawa's relation is not yet optimal. Recently, Branciard \cite{Branciard13} tightened Ozawa's EDR via
\begin{eqnarray}\label{eq:Branciard}
&& \epsilon^2(A)\Delta^2(B)+\Delta^2(A)\eta^2(B)\nonumber\\
&+&2\epsilon(A)\eta(B)\sqrt{\Delta^2(A)\Delta^2(B)-C^2_{AB}}\geq C^2_{AB}\,,
\end{eqnarray}
with $C_{AB}=\frac{1}{2}\vert\bra{\psi} [A,B] \ket{\psi}\vert$.
A tight error-disturbance uncertainty relation, describing the optimal trade - off  relation between error $\epsilon(A)$ and disturbance $\eta(B)$ is obtained in the special case of $\Delta(A)=\Delta(B)=1$ and replacing $\epsilon(A)$ and $\eta(B)$ by $\tilde{\epsilon}(A)=\epsilon(A)\sqrt{\frac{\epsilon^2(A)}{4}}$ and   $\tilde{\eta}(B)=\eta(B)\sqrt{\frac{\eta^2(B)}{4}}$, respectively.
However, apart from r.m.s errors other error measures have been introduced in recent years: Busch and co-workers proposed an approach where error  and  disturbance  are evaluated  from  the  difference  between  output  probability distributions \cite{Busch13,Busch14}. Here, reference measurements of the target observables are performed first. Then, the output statistics of the successive measurement are then compared to the output statistics of the single (reference) measurements, yielding error and disturbance. However, there continues to be some debate as to the appropriate measure of measurement (in)accuracy and of disturbance \cite{Ozawa03,OzawaPLA03,Hall04,Branciard13,Erhart12,Steinberg12,Sulyok13,Busch13,Busch13PRA,Busch14,Ringbauer14,Kaneda14,Buscemi14,Sulyok15,Ma16,Sponar17,Barchielli18,Pan19}. 
Applying the neutron polarimetric setup from \figref{fig:Uncertainty}\,(b) (left) we demonstrated the universal validity of Ozawa's new measurement uncertainty relation, as well as a violation of Heisenbergs's original error-disturbance uncertainty relation \cite{Erhart12,Sulyok13}. This first experiments of error-disturbance measurement uncertainty relations, triggered theoretical as well as experimental studies in the field.
Subsequent all-optical setups achieved similar results \cite{Steinberg12,Edamatsu13,Kaneda14,Ringbauer14,Ma16,Pan19}.
In \figref{fig:Uncertainty}\,(c) results of error and disturbance plotted as $\epsilon(A)$  versus $\eta(B)$ Blue curve: the Branciard bound as defined in \eqnref{eq:Branciard}. Blue marker: experimental results. Orange curve error and disturbance as in Ozawa's definition \eqnref{eq:Ozawa}. Green curve: bound imposed by Heisenberg's original error-disturbance relation  $\epsilon(A)\eta(B)\geq\frac{1}{2}\vert\bra{\psi} [A,B] \ket{\psi}\vert$. At this point we want to emphasize a peculiarity of Pauli Spin observables (in  qubit systems), namely the \emph{state-independence} of Ozawa's definition for error $\epsilon(A)$ and disturbance $\eta(B)$, which applies to pure, as well as to mixed states, and was demonstrated with neutrons in \cite{Sulyok13,Demirel16}. An experimental test of a \emph{state-dependent} reformulation of Ozawa's quantum rms-error for projective (but not Pauli-type) measurements \cite{Ozawa19}, in response to a counter example proposed by Busch and co-worker \cite{Busch04}, is reported in \cite{Sponar21}.

The recent interest in \emph{measurement} uncertainty relations revealed that the well-known Robertson-Schr\"{o}dinger uncertainty relation lacks an irreducible or state-independent lower bound. Consequently, Busch \emph{et al.} \cite{Busch14} proposed state-independent uncertainty relations for arbitrary Pauli observables $A=\boldsymbol a\cdot\boldsymbol\sigma$ and $B=\boldsymbol b\cdot\boldsymbol\sigma$ of a two-level system (qubit), expressed as $\Delta A+\Delta B\geq \vert\boldsymbol a\times \boldsymbol b\vert$ in 2014. 
However, this relation is not tight in general. Hence, Abbott et al., proposed a state-independent tight \emph{preparation} uncertainty expressed as  
\begin{eqnarray}\label{eq:StDec}
&\Delta^2 ( A)+\Delta^2  (B)+2\vert\boldsymbol a\cdot\boldsymbol b\vert\sqrt{1-\Delta^2 ( A)}\sqrt{1-\Delta^2 ( B)} \nonumber\\& \ge1+(\boldsymbol a\cdot\boldsymbol b)^2,
\end{eqnarray}
for standard deviation $\Delta (A)$ and $\Delta (B)$ in \cite{Abbott16}.
Applying the neutron polarimetric setup depicted in \figref{fig:Uncertainty}\,(d) \eqnref{eq:StDec} was tested recently and reported in \cite{Sponar20}.
In \figref{fig:Uncertainty}\,(e) experimental results of Abbott's tight state-independent preparation uncertainty, together with the lower bound from Busch \cite{Busch14} and the (trivial) bound from Robertson \cite{Robertson29} can be seen.

It is widely accepted \cite{Deutsch83}, that the uncertainty relation as formulated by Robertson in terms of standard deviations $\Delta(A,\vert\psi\rangle)\Delta(B,\vert\psi\rangle)\geq\frac{1}{2}\vert\langle\psi\vert[A,B]\vert\psi\rangle\vert$ lacks an irreducible or state-independent lower bound, meaning it can become zero for non-commuting observables. Furthermore, the standard deviation is not an optimal measure for all states. Consequently, Deutsch began to seek a theorem of linear algebra in the form  $\mathcal U(A, B,\psi)\ge\mathcal B( A, B)$ and suggested using (Shannon) \emph{entropy} as an appropriate measure. Note that Heisenberg's (and Kennard's) inequality $\Delta Q\Delta P\ge\frac{\hbar}{2}$ has that form, but its generalization Eq.\,(\ref{eq:Robertson}) does not. Uncertainty relations in terms of entropy were introduced to solve both problems. The first \emph{entropic uncertainty relation} was formulated by Hirschman \cite{Hirschman57} in 1957 for the position and momentum observables, which was later improved in 1975 by Beckner \cite{Beckner75}. The extension to non-degenerate observables on a finite-dimensional Hilbert space was given by Deutsch in 1983 \cite{Deutsch83} and later improved by Maassen and Uffink \cite{Maassen88} yielding the well-known \emph{entropic} uncertainty relation 
\begin{equation*} \label{eq:Deutsch83}
H( A)+H( B)\geq -2\,{\rm log}\,c, \qquad c :=\max_{i,j} |\langle a_i|b_j\rangle|,
\end{equation*}
where $H$ denotes the Shannon entropy and $c$ is the maximal overlap between the eigenvectors $\ket {a_i}$ and $\ket{ b_j}$ of the observables $A$ and $B$. In the case of qubits, standard deviations $\Delta A$ and $\Delta B$, as well as the Shannon entropy of a Pauli observable $X$, $H(X)$, can be directly expressed in terms of the expectation value $\langle  X\rangle$, namely as $(\Delta  X)^2=1-\langle  X\rangle^2$ and
$
H( X)=h_2[\frac{1+\langle X\rangle}{2}],
$
where where $h_2$ is the binary entropy function defined as 
$
h_2(p)=-p\,{\mathrm {log}}\,p-(1-p)\,{\mathrm {log}}\,(1-p).
$
Then \eqnref{eq:StDec} can be reformulated in terms of entropies for two Pauli observables $A$ and $B$ as
\begin{eqnarray*}\label{eq:Entropy}
f\big(H(A)\big)^2+f\big(H(B)\big)^2-2\vert\vec a\cdot\vec b\vert\,f\big(H(A)\big)\,f\big(H(B)\big)\nonumber\\
\leq\big(1-(\vec a\cdot\vec b)^2\big)\vert\vec r\vert^2\leq 1-(\vec a\cdot\vec b)^2,\nonumber\\
\end{eqnarray*}
with $f(x):=1-2h^{-1}_2[x]$, where   $h^{-1}_2$ denotes the inverse function of  $h_2$  (see \cite{Sponar20} for experimental results).

So it is very natural to also seek an \emph{entropic measurement uncertainty} relation, based on information gain and loss due to a measurement process. Such a formulation as recently introduced by Buscemi \emph{et al} in \cite{Buscemi14}, introducing a state-independent information-theoretical uncertainty relation. In the presented framework, noise and disturbance are quantified not by a difference between a system observable and the quantity actually measured, but by the correlations between input states and measurement outcomes. 
The noise is defined in the following scenario: the eigenstates of $A$ are randomly prepared with equal probability. If $\mathcal{M}$ accurately measures $A$ then the value of $m$ should allow one to infer $a$; if the measurement is noisy, $m$ yields less information about $a$. This noise is quantified in terms of the conditional Shannon entropy: denoting the random variables associated with $a$ and $m$ as $\mathbb{A}$ and $\mathbb{M}$, respectively, the \emph{noise} of $\mathcal{M}$ for a measurement of $A$ is~\cite{Buscemi14} 
\begin{equation*}
	N(\mathcal{M},A) = H(\mathbb{A}|\mathbb{M})=-\sum_{a,m} p(a,m)\log p(a|m),
	\label{eq:noiseDefnOrg}
\end{equation*}
where $p(a,m)=p(a)p(m|a)$ and $p(a|m)$ can be calculated from Bayes' theorem.
The disturbance $D_\mathcal E(\mathcal{M}, B)$ is defined as the conditional entropy $H(\mathbb{B}|\mathbb{B}'_{\mathcal M,\mathcal E})$ as
\begin{equation*}
D_{\mathcal{E}}(\mathcal{M}, B) :=H(\mathbb{B}|\mathbb{B}'_{\mathcal M,\mathcal E}) = -\sum_{i,j}p(b_i, b'_j) \log(p(b_i| b'_j))~.
\label{eq:DistDefn}
\end{equation*}
%
%
Using these notions of noise and disturbance, for arbitrary observables $A$ and $B$ in finite-dimensional Hilbert spaces, the noise-disturbance (measurement) relation
\begin{equation*}
	N(\mathcal{M}, A) + D_{\mathcal{E}}(\mathcal{M}, B) \geq -\log [\max_{i,j} |\langle a_i|b_j\rangle|^2]
	\label{eq:NoiseDistUR}
	\end{equation*}
holds \cite{Buscemi14}.
In \cite{Sulyok15} we experimentally tested 
\begin{equation*}
	g[N(\mathcal{M}, A) ]^2+g[ D_{\mathcal{E}}(\mathcal{M}, B)]^2 \leq 1,
	\label{eq:NoiseDistUR}
	\end{equation*}
where $g[x]$ is the inverse of the function $h(x)$ defined as
$ h(x)=-\frac{1+x}{2}\log\left(\frac{1+x}{2}\right)-\frac{1-x}{2}\log\left(\frac{1-x}{2}\right), \,x \in [0,1]$ for maximally incompatible Pauli spin observables. However, this relation for projective measurements is outperformed by \emph{general}, or more precisely \emph{positive-operator-valued-measures} (POVM). This is demonstrated in a neutron polarimetric experiment for a four-outcome POVM in \cite{Demirel19} and for a three-outcome POVM in \cite{Sponar21POVM}, respectively (see \cite{Demirel20} for a review of entropic uncertainty relations).


\section{Tests of Gravity and Non-Newtonian Interactions}
\label{sec:grav}

{\red{}Another topic in which neutrons have contributed significantly concerns the investigation of (modified) gravity as well as DM and DE.}
The theoretical basis and relevant interaction potentials are briefly surveyed in \secref{sec:grav:MGDMDE}.
Testing the plethora of DM and DE models requires a variety of different experiments. Due to their small dielectric polarizability, and vanishing electric charge and dipole moment, neutrons are not susceptible to most electrostatic background effects affecting other test particles. For this reason, neutron experiments give the tightest constraints on parameters for some models at interaction ranges between femtometers and micrometers. 
In Sections~\ref{sec:grav:scattering} -- \ref{sec:grav:other}, we briefly review the most important experimental methods used in neutron physics to perform precision tests of dark interactions. Finally, limits on the parameters of these interactions are collected in \secref{sec:grav:limits}.

\subsection{The Dark Sector -- An Overview}
\label{sec:grav:MGDMDE}

Our current knowledge of fundamental physics is based on two pillars, the standard model of elementary particle physics (SM) and Einstein's general theory of  relativity (GR). While both theories have been enormously successful in restricted domains of application, they leave many important questions unanswered. 
Despite intense efforts over several decades by some of the most distinguished physicists of the modern era, unification in a consistent quantum theory of gravitation has not yet been achieved. This is notwithstanding the progress made by string theory, loop quantum gravity and other approaches to quantum gravity. As is now known, the particle content of the SM describes only a small fraction of the matter and energy distribution in our current Universe.

As long ago as the 1930s Zwicky \cite{Zwicky:1933, Zwicky:1937zza} postulated the existence of dark matter after studying galaxy clusters and obtaining evidence of unseen mass (the prehistory of dark matter can be traced back even further to earlier investigations done by Lord Kelvin and Poincar\'e \cite{Bertone:2016nfn}). 
The existence of dark matter is by now well established through numerous studies of astronomical objects \cite{Bertone:2010zza}. However, even after decades of investigation, the origin of dark matter is still unknown; the only certainty being that this origin is not to be found within the SM.

\begin{figure*}[!t]
\begin{tikzpicture}
  \node[draw=tublue,thick,rounded corners=3pt,inner sep=\boxtextsept,text width=\innertextwidth,text justified,anchor=north]
  {\mbox{}\vspace{12pt}\\{\small
Hypothetical DM interactions {\red{}typically} involve spin-0 or spin-1 bosons. For the exchange of a spin-0 particle $\phi$ between two fermions $\psi$, we have the {\red{}interaction} Lagrangian
\begin{align}
 \mathcal{L}_\phi=\phi\sum\limits_\psi \bar{\psi}\left(g_{S,\psi}+\ri\gamma^5g_{P,\psi}\right)\psi\,,\label{eq:grav:lagr_spin0}
\end{align}
with the coupling constants $g_{S,\psi}$ for scalar and {\red{}$g_{P,\psi}$} for pseudoscalar interactions and the Dirac matrix $\gamma^5=i\gamma^0\gamma^1\gamma^2\gamma^3$. \eqnref{eq:grav:lagr_spin0} leads to the potentials~\cite{Fadeev:2018rfl}
\begin{align}
V_{SS}(r)&=-\frac{g_{S,1}g_{S,2}\hbar c}{4 \pi}\frac{\re^{-r/\lambda}}{r}\quad &&\text{scalar}\label{eq:scalar}\\
 V_{SP}(r)&=-\frac{g_{S,1} g_{P,2} \hbar}{4\pi m_2}\,\bm{s}_2\cdot\hat{\bf{r}}\,\left(\inv{r}+\inv{\lambda}\right)\,\frac{\re^{-r/\lambda}}{r}+(1\leftrightarrow2) &&\text{scalar-pseudoscalar}\label{eq:scalar-pseudoscalar}
 \end{align}
depending on separation $r$ between fermions 1 and 2 with unit vector $\hat{\bf{r}} = {\bf{r}}/r$, (neutron) spin $\bm{s}=(\hbar/2)\bm{\sigma}$, and interaction range $\lambda$. Permutation symmetry requires the addition of the same potential with exchanged indices $(1\leftrightarrow 2)$ {\red{}for the potentials arising from two different vertices}. Exchange of a {\red{}massive} spin-1 boson $Z'$ is given by
\begin{align}
 {\red{}\mathcal{L}_{Z'}=Z'_\mu\sum\limits_\psi \bar{\psi}\gamma^\mu\left(g_{V,\psi}+\gamma^5g_{A,\psi}\right)\psi}\,,\label{eq:grav:lagr_spin1}
\end{align} 
and the position-space interaction potentials
\begin{align}
V_{AV}(r)&=\underbrace{\frac{g_{A,1}g_{V,2}}{4\pi}\Bigg[\bm{s}_1\hspace{-1.5pt}\cdot\hspace{-1.5pt}\left\{\frac{\bm{p}_1}{m_1}\hspace{-1pt}-\hspace{-1pt}\frac{\bm{p}_2}{m_2},\frac{\re^{-r/\lambda}}{r}\right\}}_{\tilde{V}_{AV}(r)}-\frac{2}{m_2}(\bm{s}_1\hspace{-1pt}\times\hspace{-1pt}\bm{s}_2)\cdot\hat{\bm{r}}\hspace{-1pt}\left(\inv{r}+\inv{\lambda}\right)\hspace{-1pt}\frac{\re^{-r/\lambda}}{r}\Bigg]\hspace{-1.5pt}+(1\leftrightarrow2)\ &&\text{vector-axial vector}\,\label{eq:vector-axialvector}\\
\tilde{V}_{AA}(r)&=\frac{\tilde{g}_A^2 \hbar}{8\pi m_n c}\bm{s}\cdot\left(\mathbf{v}\times\hat{\bf{r}}\right)\left(\inv{r}+\inv{\lambda}\right)\frac{\re^{-r/\lambda}}{r}&&\text{dual axial vector}\label{eq:axialvector}
\end{align}
where $\bm{p}_i$ are the configuration space momentum operators and $\{,\}$ denotes the anti-commutator.
{\red{}We note that in neutron experiments only limits for the potentials $\tilde{V}_{VA}$ and $\tilde{V}_{AA}$~\cite{Dobrescu:2006au} have been derived. It has been pointed out~\cite{Fadeev:2018rfl} that the replacement of the operator $\bm{p}/m$ by the classical vector $\bm{v}$ is only correct for macroscopic masses, but not for atomic or sub-atomic particles, and that some expressions for potentials in Ref.~\cite{Dobrescu:2006au} are incomplete. Note further that $\tilde{V}_{AA}$ is a term of higher order in $v/c$, while the zero-order potential $V_{AA}$ contains terms $\propto \bm{s}_1\cdot\bm{s}_1$ and $(\bm{s}_1\cdot\hat{\bm{r}})(\bm{s}_2\cdot\hat{\bm{r}})$ that vanish if one of the interacting bodies is unpolarized. Also \eqnref{eq:grav:lagr_spin1} leads to a double-vector potential $\tilde{V}_{VV}$ of identical operator structure as $\tilde{V}_{AA}$ that cannot be considered independently of the latter (see Eq.(5.28) of Ref.~\cite{Dobrescu:2006au}).}
Expressions for the remaining potentials $V_{PP}$, $V_{VV}$, and massless spin-1 boson exchange can be found in Ref.~\cite{Fadeev:2018rfl}, but have not been considered in neutron experiments. Historically, generic Yukawa interactions have directly been related to modifications of the Newtonian potential with the identification {\red{}$\a=-g_S^2 \hbar c/(4 \pi G_N m_n m_2)$}, where either $m_2=m_n$ for neutron scattering and $m_2=m_\text{Earth}$ for most other experiments.\\*
We also give limits for the screened quintessence models chameleon and symmetron. {\red{}For low densities }they are given by the Lagrangians~\cite{Joyce:2014kja}
\begin{align*}
 \mathcal{L}_\text{Cha}&=\inv{2}\,\partial_\mu\phi\,\partial^\mu\phi - \frac{\Lambda^{4+n}}{\phi^n} - \rho\,\frac{\beta\,\phi}{M_\text{Pl}}\,, \\
 \mathcal{L}_\text{Sym}&=\inv{2}\,\partial_\mu\phi\,\partial^\mu\phi - \inv{2}\left(\frac{\rho}{M^2} - \mu^2\right)\phi^2 - \frac{\lambda}{4}\,\phi^4\,,
\end{align*}
where $\phi$ denotes the scalar field, $\Lambda$, $\beta$, $M$, and $\mu$ are model parameters, $\rho$ is the mass density of the environment, and $M_\text{Pl} = 1/\sqrt{8\pi G_N}$ is the reduced Planck mass. Depending on the experimental setup, the scalar field takes on a field profile and induces a potential $V(r)$.
}};
\node[draw=tublue,thick,inner sep=3pt,text width=\innertextwidth+14pt,text justified,anchor=north,fill=tublue!20] {\vspace{-11pt}
\mbox{}\hspace{3pt}\begin{minipage}{\innertextwidth}
\vspace{2pt}
 \subsubsection*{Box 3: Potentials of Hypothetical Dark Interactions}
\vspace{-6pt}\mbox{}
\end{minipage}
};
\end{tikzpicture}
\end{figure*}

Possible candidates for dark matter have been found in hypothetical particles that were devised to tackle unresolved problems of the SM.
In order to solve the so-called strong CP problem, Peccei and Quinn proposed an elegant mechanism \cite{Peccei:1977hh, Peccei:1977ur}, which 
contained a hypothetical new pseudoscalar particle, termed axion~\cite{Weinberg:1977ma, Wilczek:1977pj, Kim:1979if, Shifman:1979if, Dine:1981rt}. The strong CP problem concerns the existence of a CP violating term in the QCD Lagrangian, which is proportional to a parameter $\theta$. This term induces electric dipole moments (EDMs), e.g. for the neutron. On dimensional grounds, one would expect this parameter to be of the order of  $\theta\sim\mathcal O(1)$. However, $\theta$ receives additional contributions from the electro-weak sector as well as from radiative corrections, which together contribute to an effective $\bar\theta$. The strong limits on the neutron EDM constrain this parameter to a significantly smaller value $\bar\theta\lesssim 10^{-10}$. Hence, the question arises as to why several contributions, unrelated in the SM, all combine to a total contribution, which is several orders of magnitude smaller than its estimated individual contributions. This fine-tuning problem bears some resemblance to the cosmological constant problem discussed below. 
In the Peccei-Quinn mechanism, a further contribution to $\bar\theta$ comes from a dynamical field, the quantum of which is the (QCD) axion. Thereby, the $\bar\theta$ parameter is effectively promoted to a dynamical field, whose dynamical relaxation in the axion potential solves the strong CP problem.
The underlying physics is related to the Higgs mechanism of the SM, with the axion arising as the "pseudo-Nambu-Goldstone" boson through spontaneous symmetry breaking of a global continuous symmetry in QCD. 
From robust astrophysical constraints based on the observation of the neutrino signal from supernova 1987A and from star cooling, we know that the axion mass has to be very small ($\lesssim10$ meV \cite{Raffelt:1999tx}). This implies long-range forces, which are in principle observable in laboratory experiments \cite{Moody:1984ba}. Heavier axions can be ruled out, as they would have produced observable effects in astrophysical objects or terrestrial experiments. On the other hand, lighter axions are not ruled out experimentally and even have a rather strong theoretical motivation \cite{Freivogel:2008qc, Linde:1987bx}.
The search for axions has obtained significant impetus, when it was realized that they provide a possible dark matter candidate \cite{Bertone:2010zza, Raffelt:1999tx, Duffy:2009ig, Graham:2015ouw}.

A plethora of other proposals for light bosons are collectively called axionlike particles (ALPs). Among them are familons, which are the "pseudo-Nambu-Goldstone" bosons related to the spontaneous breaking of flavor symmetry \cite{Davidson:1981zd, Wilczek:1982rv, Gelmini:1982zz}, majorons, which were proposed in order to understand neutrino masses \cite{Chikashige:1980ui, Gelmini:1980re} and arions, which are the bosons related to the spontaneous breaking of chiral lepton symmetry \cite{Anselm:1982}. Furthermore, new spin-0 or spin-1 gravitons have been hypothesized \cite{Scherk:1979aj, Neville:1980dd, Neville:1981ut, Carroll:1994dq}. 
Axion-like particles arise in string theory as excitations of quantum fields extending into compactified extra space-time dimensions \cite{Bailin:1987jd, Svrcek:2006yi}. A further proposition in the context of string theory proposes a so-called axiverse, the existence of many ultralight ALPs \cite{Arvanitaki:2009fg}.
Furthermore, axions and ALPs have even been proposed as a possible solution to the hierarchy problem, which, in a nutshell, concerns the question of why the Higgs mass is so much lighter than the Planck mass, contrary to expectations \cite{Graham:2015cka}.
In order to distinguish QCD axions from generic ALPs the experimental results of nuclear EDM searches have been compared to those from direct searches in fifth-force experiments and from combining laboratory searches with astrophysical bounds on stellar energy loss \cite{Mantry:2014zsa, Mantry:2014} (see also \cite{Gharibnejad:2014kda}). 

In addition to spin-0 bosons, a host of hypothetical spin-1 particles related to new $U(1)$ gauge symmetries have been proposed. Unbroken $U(1)$ gauge symmetries corresponding to massless bosons arise naturally in string theory and other standard model extensions \cite{Cvetic:1995rj}. Those particles are generically termed paraphotons \cite{Holdom:1985ag}. Further related proposals are those of the dark photon \cite{Ackerman:2009} and $Z'$ bosons. The latter are related to the $Z$ boson of the SM and arise in numerous theoretical models with broadly varying theoretically motivated masses and couplings to quarks and leptons \cite{Langacker:2008yv}.

Each of these hypothetical new particles induces non-relativistic potentials, thereby leading to experimentally observable effects \cite{Dobrescu:2006au, Fadeev:2018rfl}. We present a selection of these potentials that have been constrained by experiments with neutrons in Box 3.

A more recent discovery concerns studies on the expansion of our universe. Investigations on type Ia supernovae revealed that the expansion of the universe is currently accelerating, contrary to expectations \cite{Perlmutter:2012, Riess:2012, Schmidt:2012xxa}. The standard theory of cosmology describing the evolution of the universe, GR, cannot account for the accelerated expansion in its original formulation. A viable solution for this problem is the inclusion of a so-called cosmological constant $\Lambda$, an extension of the theory already devised by Einstein in 1917 \cite{Einstein:1917}. Doing so, however, leads to a severe fine-tuning problem, since in addition to Einstein's original (bare) cosmological constant, there are contributions from the zero-point energies of all particles of the SM as well as others that are still unknown. Furthermore, the Higgs potential leads to an effective contribution to the total cosmological constant during the phase transition related to the electro-weak symmetry breaking. All of these contributions have to add up to the very small measured value of the total cosmological constant $\Lambda\sim10^{-52}\,{\rm m}^{-2}$. Theoretical estimates, together with experimental observations, show that the contributions of the zero-point energies of SM particles are some 55 orders of magnitude larger than $\Lambda$ itself, amounting to incredible fine-tuning \cite{Sola:2013gha}. While GR itself may easily be modified at short distance scales by inclusion of higher derivative terms, it proves to be a remarkably rigid theoretical construction at larger distance scales. In fact, modifications at cosmological scales often lead to theoretical inconsistencies. Consequently, instead of modifying GR itself, it seems more natural to add hypothetical new fields to solve the fine-tuning problem. This leads to the postulation of an as-yet unknown substance, so-called dark energy. Similarly, as for dark matter, the origin of dark energy is so far unknown.

The Planck Mission study of the cosmic microwave background radiation \cite{Adam:2015rua} revealed that the present universe consists of 69\% dark energy, 26\% dark matter and only 5\% SM matter. Consequently, 95\% of the current matter/energy content of our universe lies `in the dark'. This is matched by a plethora of new theory proposals that have been devised to account for some of the vexing questions left unanswered by the SM as well as GR.

Quintessence denotes a proposal to describe the accelerated expansion of the universe by mimicking the effect of a time-dependent cosmological `constant' via the potential energy of a hypothetical new scalar field (for reviews see \cite{Padmanabhan:2002ji, Peebles:2002gy, Frieman:2008sn, Linder:2007wa, Tsujikawa:2013fta, Joyce:2014kja}). The existence of new scalar fields would generically also imply new interactions, so-called fifth forces. These, however, are strongly constrained at distance scales of the solar system and below \cite{Will:2014kxa} by manifold observational tests. Various screening mechanisms have been devised to avoid those observational constraints. Screening mechanisms suppress scalar fields or their interaction with SM matter in observational regions where the matter density is high, while they allow the scalar field to prevail in interstellar space, where it can effectively drive the cosmic expansion. 
One class of proposed screening mechanisms relies on non-linear self-interaction terms in the effective potential, such that the latter depends on the energy density of the local environment. In this way, the field profile reacts to its surroundings. In this class of screening mechanisms, one distinguishes three ways in which a scalar field can be suppressed in dense environments \cite{Burrage:2017qrf}. Firstly, the effective mass becomes larger, i.e. the corresponding interaction range becomes very short, secondly, the coupling to matter becomes small, and/or thirdly, not all of the mass sources the field. So-called chameleon models utilize a combination of the first and third condition \cite{Khoury:2003rn, Khoury:2003aq, Mota:2006ed, Mota:2006fz, Waterhouse:2006wv}. By now, chameleons have been extensively searched for in several laboratory experiments. Another model that utilizes the second condition is known as the symmetron model, which utilizes an effective potential similar to the Higgs potential \cite{Hinterbichler:2010es, Hinterbichler:2011ca, Pietroni:2005pv, Olive:2007aj}. Some technical details on chameleons and symmetrons are provided in Box 3. Yet another model utilizing the second condition is the dilaton model, which is of special interest due to its string theory motivation. It was first predicted in this context by Damour and Polyakov \cite{Damour:1994zq} (see also \cite{Brax:2010gi, Gasperini:2001pc}). If it is combined with an environment-dependent Damour-Polyakov coupling, it leads to a well-motivated screened dark energy model. Unfortunately, these scalar fields still require a certain amount of fine-tuning as well as the addition of a non-zero cosmological constant as introduced by Einstein. Nevertheless, they offer the simplest extension of the cosmological standard model \cite{Joyce:2014kja} and are therefore of interest.

A second class of proposed screening mechanisms relies on non-linearities in the kinetic sector and screen through what is known as the Vainshtein mechanism \cite{Vainshtein:1972sx, Babichev:2013usa, Joyce:2014kja}. Although of great theoretical interest, Vainshtein-screened models generically cannot be probed by laboratory experiments.
Eventually, an approach that stands independent of quintessence or screening is known as modified Newtonian dynamics (MOND) \cite{Milgrom:1983ca}. 
Here, Newtonian gravity is modified at long distances. A representation of DE as a ghost condensate has been proposed in Ref.~\cite{ArkaniHamed:2003uy}. This ghost condensate, a constant-velocity scalar field, acts as a fluid filling the universe and mimics a cosmological constant due to its negative kinetic energy term. The interaction of the ghost condensate with SM matter would lead to several observable effects, i.e. Lorentz-violating effects as well as new long-range spin-dependent interactions \cite{ArkaniHamed:2003uy, ArkaniHamed:2004ar}. 

Both dark matter and dark energy models can be excluded in laboratory experiments sensitive to the respective interaction potentials. As neutrons are massive and have spin, they provide a sensitive and versatile tool for probing such potentials~\cite{Abele:2008zz,Brax:2011hb} (see Box 3 for further details). Neutrons therefore provide access to the dark sector complimentary to modern colliders, which are becoming increasingly expensive and may reach a saturation with respect to particle energies and intensities in the foreseeable future.

\subsection{Neutron Scattering}
\label{sec:grav:scattering}
Neutron scattering (reviews~\cite{Rauch:2000,Willis:2009}) on solid or gaseous targets is a well-established method to not only measure scattering parameters but also to derive limits on non-Newtonian forces. In the experiment, thermal neutrons with wavelengths of a few $\si{\angstrom}$ and momentum $k_{in}$ are directed onto a solid or fluid target where they interact with the nuclei of the material. Since these nuclei have fm diameters, the interaction is well described in the Born approximation as point-like. Accordingly, the isotropically or Bragg (s-wave) scattered intensity $I(q)$ is recorded by large-scale detectors and thus depends exclusively on the difference $\mathbf{q}=\mathbf{k}_{out}-\mathbf{k}_{in}$ by which the neutron momentum is changed in the interaction. The independence of the geometric angle renders neutron scattering robust, as $q$ can be determined conveniently from time-of-flight information.\\
\begin{table*}[t]
\centering
\caption{Values of the various contributions to the scattering lengths in [fm] for commonly used scatterers and unpolarized neutrons at $q=\SI{2}{\nano\metre^{-1}}$. For details of the terms and calculations see the review by Sears~\cite{Sears:1986}. We use charge radii from Ref.~\cite{Angeli:2013epw}. The Yukawa scattering length is given for $\lambda=\SI{1}{\nano\metre}$.\label{tab:scat_len}}
 \begin{tabular}{l c r@{.}l r@{.}l r@{.}l r@{.}l r@{.}l}
 Material & $Z$ & \multicolumn{2}{c}{$b_N$} & \multicolumn{2}{c}{$b_M$} &\multicolumn{2}{c}{$b_E$} & \multicolumn{2}{c}{$b_P$} & \multicolumn{2}{c}{$b_Y/\a$}\\
 \hline
 $^{208}$Pb & 82 & 9&49 & 0&113 & -0&110 & 0&047 & \multicolumn{2}{r}{$2.32\times10^{-24}$}\\
 Xe (natural) & 54 & 4&69 & 0&075 & 0&073 & 0&024 & \multicolumn{2}{r}{$1.12\times10^{-26}$}\\
\end{tabular}
\end{table*}
Apart from a factor $C_{\text{exp}}$ determined by the experimental setup~\cite{Haddock:2017wav}, the intensity can be calculated exactly from~\cite{Voronin:2018rmz,Pokotilovski:2006up,Rauch:2000} 
 Apart from a factor $C_{\text{exp}}$ determined by the experimental setup~\cite{Haddock:2017wav}, the intensity can be calculated exactly from~\cite{Voronin:2018rmz,Pokotilovski:2006up,Rauch:2000} 
 \begin{align}
  I(q)=C_{\text{exp}}[b(q) S]^2\,,\label{eq:grav:nscattering:intensity}
 \end{align} 
\begin{align}
b(q)=\frac{m_n}{2\pi\hbar^2}\int{\rm d}\mathbf{r}\, V(r)\re^{\ri \mathbf{q}\mathbf{r}}\,,\label{eq:grav:nscattering:slen}
\end{align}
depending on the interaction potential $V(r)$ between a neutron and an atom at distance $r$. Now $b$ [or respectively $V(r)$] can be separated into different contributions described in detail in the literature~\cite{Sears:1986}: Firstly, the nuclear interaction $b_N(q)$ can be approximated by a Fermi delta-potential in the slow neutron regime $k_{in}r\ll1$. Secondly, the neutron's magnetic dipole moment and spin give rise to electromagnetic interactions $b_M(q)$ with the field and spin of the nucleus. Thirdly, 
the neutron contains the distributed charges of its quarks, leading to a non-vanishing electric polarizability. The associated electrostatic $b_E(q)$ and polarization $b_P(q)$ terms are at least two orders of magnitude smaller than nuclear and magnetic contributions, for which they are mostly neglected in experimental evaluations. For measurements with unpolarized neutrons or nuclei, the magnetic term is also strongly reduced. In practice, the different contributions are rewritten in the form  
\begin{align}
 b(q)=b_c(q)\hspace{-1pt}+b_i(q)[I(I\hspace{-1pt}+\hspace{-1pt}1)]^{-1/2}\bm{\sigma}\cdot\mathbf{I}+\text{magnetic terms}\nonumber,
\end{align}
with the coherent and incoherent scattering lengths $b_c(q)$ and $b_i(q)$, respectively. Additional minor corrections are discussed in the literature~\cite{Sears:1986}. 
Experimentally, the total scattering length is determined from the intensity via \eqnref{eq:grav:nscattering:intensity}. For Bragg diffraction, $q$ is fixed by the scattering angle $2\theta$ to $q=4\pi/\lambda\,\sin\theta$, while for the transmission method $q$ can be varied over a wide range. Theoretical and experimental accuracy in the determination of $b(q)$ both are currently at the $10^{-3}$--$10^{-4}$ level for which hardly any improvement is expected in the near future, although, for some materials more precise measurements have been achieved~\cite{Abbas:2012}. If we add a Yukawa-type potential $V_Y(r)$ (or $V_{SS}$) in \eqnref{eq:grav:nscattering:slen}, we obtain another contribution
\begin{equation}
b_Y(q)=2\a \frac{G_N m_n^2 M_A}{\hbar^2}\frac{\lambda^2}{1+(q\lambda)^2}\,,\label{eq:grav:nscattering:bYukawa}
\end{equation}
which would add to $b_c(q)$. Limits on the interaction strength $\a$ (see Box 3) can thus be obtained from the non-observation of a difference between experimental data and theoretical predictions. For typical values of $q$ around $(1\,\si{\angstrom})^{-1}$, scattering experiments can only be sensitive at very short separations $\lambda\lesssim \SI{1}{\nano\metre}$. In order to obtain a better understanding for the size of the mentioned effects, we consider Pb and Xe as typical examples. As is obvious from \tabref{tab:scat_len}, only values $\a\gtrsim10^{25}$ can be detected in the experiment. 

 The first limits on non-Newtonian forces on the basis of $^{208}$Pb neutron scattering data were reported by Leeb and Schmiedmayer~\cite{Leeb:1992qf}. 
A more detailed analysis of neutron scattering data from $^{208}$Pb in the \SI{1.26}{\electronvolt}--\SI{24}{\kilo\electronvolt} range with respect to limits on dark interactions was reported by Pokotilovski~\cite{Pokotilovski:2006up}. Nesvizhevsky \emph{et al}~\cite{Nesvizhevsky:2007by} combined data from 13 independent measurements of forward and total crosssections to obtain limits $\a q^2\l^4/[1+(q\l)^2]\leq0.0013\,{\rm fm}^2$ in the range \SI{1}{\pico\metre}--\SI{5}{\nano\metre}. Voronin \emph{et al}~\cite{Voronin:2018rmz} performed Bragg scattering on polycrystalline silicon and extracted limits from their data. The best current limits on general Yukawa forces from neutron scattering have been obtained using a gaseous Xe target~\cite{Kamiya:2015eva,Haddock:2017wav}.\\
Neutron scattering offers insight into a range of parameters among which the coherent scattering length is only one. The downside is that not only do all four fundamental forces contribute to the measured intensity, but also the dependence on $q$ needs to be modelled accurately. The low-energy tails of resonances at higher energies need to be known and taken into account via the Breit-Wigner formalism, which further complicates data evaluation. For this reason, more direct optical methods have gained popularity and are discussed below. Neutron interferometry is described separately in \secref{sec:neutron_if}. 


\subsection{Neutron Optics}
\label{sec:grav:optics}
There are several different methods, in this category that commonly use neutrons at energies $<\SI{1}{\electronvolt}$. Such neutrons have wavelengths larger than the interatomic spacings and mainly interact with the optical (average pseudo-Fermi) potential $V_F=(2\pi\hbar^2/m_n)b_c\rho_N$ of the material~\cite{Leeb:1992qf}. Here $\rho_N$ denotes the atomic density and $b_c=b_N+b_P+b_E Z[1-f(q)]$ receives contributions from neutron-nucleon interactions $b_N$, electric polarizability $b_P$, and interaction with the distributed charge of the nucleus $b_E$. The latter is scaled by the charge number $Z$ of the nucleus and the form factor $f(q)$~\cite{Sears:1986}. An extensive review of the relevant theory was given by Sears~\cite{Sears:1982}. Descriptions of the experimental methods total reflection, Christiansen-filter, and prism refraction can be found in Refs.~\cite{Sears:1982,Rauch:2000}. Neutron gravity refractometry (NGR)~\cite{Maier-Leibnitz:1962,Koester:1965} is one of the most precise methods for coherent scattering length measurements. Cold neutrons are admitted to freely fall down a well-defined height $h$ onto a target surface where they are reflected from $V_F$. For heights larger than the critical height $h_c$ the reflected intensity rapidly declines and the scattering length is extracted from $V_F=m_n g h_c$. An additional hypothetical long-range interaction would be indistinguishable from $V_F$ and therefore influence respective measurements.

One major difference between results of neutron scattering and neutron optics is that the former depend on the scattering length at $q\neq 0$ while for the latter $q=0$ is most important. Neutron optical effects are described in the same way as their counter parts in photon optics via a refractive index
\begin{align}
 n^2=\left(\frac{k_{in}}{k_{out}}\right)^2=1-\frac{4\pi\rho_N}{k_{in}^2}b_c(0)\zeta\nonumber\,,
\end{align}
where $\zeta$ is a correction factor for multiple scattering and local field effects~\cite{Warner:1985,Sears:1985} at the $10^{-4}$ level. The reason why $n$ only depends on $b_c$ at $q=0$ is that in diffraction (transmission) theory~\cite{Sears:1982} the transmitted wave is created by coherence between the incident $k_{in}$ and refracted wave $nk_{in}$ in the material, which only gives a contribution around $|n-1|\ll1$~\cite{Goldberger:1947}. At $q=0$, also most form factors reduce to unity, which simplifies the analysis and reduces uncertainties.\\
A long-range Yukawa interaction with a $q$-dependence described by \eqnref{eq:grav:nscattering:bYukawa} would influence $b_{c,o}$ measured in neutron optics at $q=0$ differently than $b_{c,s}$ determined in scattering experiments at $q\neq0$. The various experiments on $b_c$ are thus differently sensitive to hypothetical interactions.
A limit on $\a$ can be extracted by demanding $V_{Y}<(2\pi\hbar^2/m_n)(b_{c,o}-b_{c,s})$~\cite{Leeb:1992qf,Abele:2008zz} from neutron-nucleon and neutron electron scattering data~\cite{Schmiedmayer:1991zz}. Statistically, it is advantageous to combine results from different measurements~\cite{Nesvizhevsky:2007by}. 
%

\begin{figure*}[t]
\begin{tikzpicture}
  \node[draw=tublue,thick,rounded corners=3pt,inner sep=\boxtextsept,text width=\innertextwidth,text justified,anchor=north]
  {\mbox{}\vspace{12pt}\\{\small
The quantum bouncer (QB) can be described by a one-dimensional Schr\"odinger equation in vertical $z$-direction as described in Box 1, where we now have three different potentials
\begin{align}
 V(z)=\underbrace{\rule{0pt}{-8pt}m_n g z(t)}_{\text{Gravity}}\hspace{4pt} +\hspace{-4pt}\underbrace{\rule{0pt}{-8pt}V_{\rm F}\,\Theta(z-z_m)}_{\text{Fermi-pseudopotential}}+\underbrace{\rule{0pt}{-8pt}V_{\text{hyp}}(z)}_{\parbox{1.5cm}{\centering\footnotesize{hypothetical\\*potential}}}\hspace{-6pt}.\label{eq:grav:qbb:potentials}
\end{align}

 \begin{center}
 \includegraphics[scale=0.8]{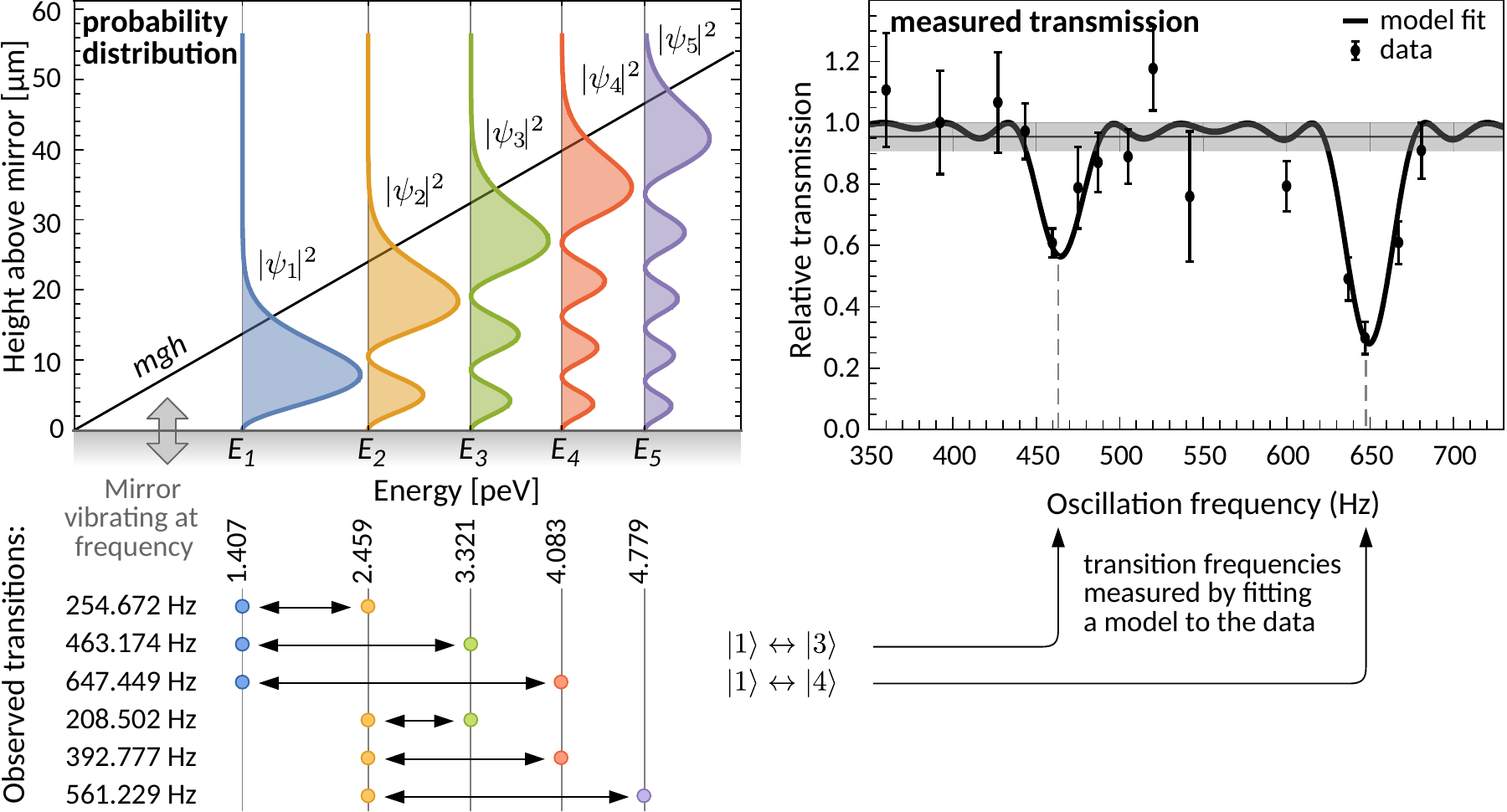}
 \caption{Quantum bouncer and GRS. Left: Lowest energy states and wave functions, right: Exemplary transmission spectrum measured in the \qbounce{} setup in Rabi configuration~\cite{Cronenberg:2018qxf}.\label{fig:grav:qbb}}
\end{center}
For $V_{\text{hyp}}(z)=0$, these potentials permit analytical solutions for the eigenstates $\psi(z,t)=\psi_n(z)\exp(-\ri/\hbar E_n t)$ with non-equidistant energies $E_n$ (see \figref{fig:grav:qbb}). Characteristically for the quantum bouncer, the spatial solutions take the form of Airy functions $\psi_{n}(z)=N_n{\rm Ai}\left[z/{z_0}-E_n/E_0\right]$ with normalization constant $N$. $z(t)$ represents the height above the resting or vibrating lower mirror.
The natural scalings of the problem $E_0=\left[\hbar^2 m_n g^2/2\right]^{1/3}=0.602\,$peV and $z_0=\left[{\hbar^2}/({2m_n^2g})\right]^{1/3}=\SI{5.87}{\micro\metre}$ already indicate that UCNs have macroscopic extensions. Since the probability distributions of states shown in \figref{fig:grav:qbb} have different vertical extensions, a hypothetical $V_{\text{hyp}}(z)$, would shift the energy of each state $|n\rangle$ --- and thereby the transition frequencies --- differently by an amount $\Delta E_n=\langle n|V_{\rm{hyp}}(z)|n\rangle$. Details on the computation of the transmission probability as a function of excitation frequency can be found in~\cite{Sedmik:2019twj,Cronenberg:2018qxf}.
}};
\node[draw=tublue,thick,inner sep=3pt,text width=\innertextwidth+14pt,text justified,anchor=north,fill=tublue!20] {\vspace{-11pt}
\mbox{}\hspace{3pt}\begin{minipage}{\innertextwidth}
\vspace{2pt}
 \subsubsection*{Box 4: Gravity Resonance Spectroscopy}
\vspace{-6pt}\mbox{}
\end{minipage}
};
\end{tikzpicture}
\end{figure*}

\subsection{Relativity \& Inertia in Neutron Interferometry}
\label{sec:grav:if} 
Early in the history of neutron interferometry it was recognized~\cite{Overhauser:1974,Colella:1975dq} that the vertical difference $\Delta h$ between the two paths of a neutron interferometer leads to a measurable phase shift $\Delta\phi_G$. By rotating the interferometer around the neutron axis, a precise differential measurement of this phase shift was implemented. However, results of this first experiment only agreed with the theoretical prediction at 88\,\%. Most of the deviation was attributed to bending of the interferometer crystal under its own weight during rotation. Soon after, Werner and co-workers~\cite{Staudenmann80} improved the measurement. The authors considered the simultaneous effects of gravity, inertia, and the quantum mechanical propagation of neutrons, leading to a remaining deviation of 4\,\%, which was later reduced further to 1\% \cite{Littrell:1997zz}. An alternative to single-crystal neutron interferometers are grating interferometers operating with very-cold-neutrons (VCN) instead of thermal neutrons (see Box 1). The longer wavelength results in a higher sensitivity and less influence of deformation during the rotation. However, despite these obvious merits, a measurement of the gravitational phase by van der Zouw~\cite{Zeilinger00} using a grating interferometer still deviated about 1\% from theory. It was suggested that the difference could be explained by a spin-dependent interaction described by $\tilde{V}_{VA}$ (see Box 3) but the conjecture was rejected by experiment~\cite{Parnell:2020wwb}.


Neutron interferometry is sensitive to any gravity-like interaction. A prime example are scalar chameleon fields, which were investigated in Ref.~\cite{Lemmel:2015kwa} as suggested by~\cite{Brax:2013cfa}. Constraints on the interaction parameter $\beta$ were derived from the vanishing phase shift difference between measurements in vacuum and in He. In the experiment~\cite{Lemmel:2015kwa} performed with unpolarized neutrons, a method originally developed for neutron scattering length measurements was employed. A chamber with different compartments for vacuum and He was inserted into the interferometer. By shifting the chamber either beam was allowed to pass through the gas or vacuum, thereby implementing a differential measurement.
\emph{Inertia}, which is a key property in both general relativity and quantum theory, can be probed in neutron interferometry using spin polarized neutrons. While inertia is usually associated with mass, intrinsic spin can exhibit similar behavior in the form of a spin-rotation coupling. This coupling is a quantum mechanical extension of the Sagnac effect resulting from the combination of the orbital angular momentum with intrinsic spin. The added spin component of the total angular momentum manifests in an additional phase of the neutron wave function \cite{Mashhoon88}. The effect can be tested by rotating the neutron spin orientation in one interferometer path by $2\pi$ using a rotating magnetic field. As the orientations before and after the manipulation are the same, only a phase difference between the two paths is induced. This phase difference results in a shift of the observed interference fringes which turns out to depend solely on the frequency of the rotation of the magnetic field \cite{Danner19,Danner20}.

\subsection{The Quantum Bouncing Ball and Gravity Resonance Spectroscopy}
\label{sec:grav:grs}

Already predicted in 1959~\cite{Zel'dovic:1959}, ultracold neutrons (UCNs, see Box 1) with energies less than $300\,$ neV 
 were found in 1969~\cite{Lushchikov:1969,Steyerl:1969csa}. Similarly as in neutron optics (discussed in \secref{sec:grav:optics}), UCNs do not interact with single nuclei or atoms, but with the collective Fermi pseudopotential $V_F$ of a material surface. As $V_F$ is typically larger than \SI{100}{\nano\electronvolt} for most materials, UCN can reflect specularly from surfaces and even be stored in `bottles'~\cite{Zel'dovic:1959}. On horizontal surfaces, UCNs form bound states in the potential well created by $V_F$ below and the (in good approximation locally linear) gravitational potential -- a system known as quantum bouncer (QB)~\cite{Gibbs:1975} or quantum bouncing ball (QBB). 
The neutron QB has been investigated experimentally and theoretically~\cite{Gibbs:1975,Gea-Banacloche:1999}. First experimental evidence for the existence of the quantized solutions of \eqnref{eq:schroedinger_generic} with the potential \eqref{eq:grav:qbb:potentials} was given by Nesvizhevsky et al~\cite{Nesvizhevsky:2002ef} and analyzed in detail~\cite{Nesvizhevsky:2003ww,Nesvizhevsky:2005ss,Westphal:2006dj}.
From these data limits on non-Newtonian interactions were derived~\cite{Abele:2003ga,Nesvizhevsky:2004qb}. The analysis~\cite{Nesvizhevsky:2004qb} was re-evaluated later~\cite{Zimmer:2006nc}. Subsequently, the development of a new technique called Gravity Resonance Spectroscopy~\cite{Jenke:2011zz} (GRS, a nomenclature coined in analogy to magnetic resonance spectroscopy as adopted by EDM experiments~\cite{Abel:2020gbr}) proved to be the key to more sensitive measurements utilizing the non-equidistant eigenstates of the QB in modern spectroscopic methods. The strength of GRS is that it does not rely on electromagnetic interactions. Using neutrons as test particles offers the advantage of bypassing the electromagnetic background induced by van der Waals and Casimir forces and other polarizability effects.
The peV differences between the eigenstate's energies correspond to acoustic frequencies. State transitions were demonstrated using controlled mechanical oscillations of the lower boundary (mirror)~\cite{Jenke:2011zz}. Such mechanical excitations offer the advantage that magnetic gradients can be avoided altogether.\\ 
The \qbounce{} collaboration has demonstrated Rabi spectroscopy using purely mechanical excitations~\cite{Jenke:2011zz}. In their setup, UCN are velocity-selected using an aperture system before entering a sandwich structure consisting of two glass mirrors separated by a precisely calibrated gap of $\sim\SI{25}{\micro\metre}$. The upper mirror is roughened, leading to chaotic scattering of higher states. Only the lowest few states (see Fig.\,5)  are allowed to pass. In a second `region' consisting of a flat mirror, controlled vibrations at variable frequency are used to induce transitions from the initial state $|i\rangle$ to the target state $|t\rangle$. A third region, similar to the first one, filters neutrons in state $|t\rangle$ such that the detector placed after this second filter only detects neutrons \emph{not} excited by the vibration. The measured neutron transmission rate $T(\omega)$ as a function of the excitation frequency thus shows clear dips at the transition frequencies (see box 4). If an additional $z-$dependent potential $V_\text{hyp}(z)$ (in \eqnref{eq:schroedinger_generic}) existed, it would shift the energy of the each state differently and hence also shift the transition frequencies. By not detecting such a shift, limits can be set on the maximum strength of $V_\text{hyp}(z)$ (see \secref{sec:grav:limits}).\\
The GRANIT experiment~\cite{Kreuz:2009ky} intends to excite resonant transitions magnetically between quantum states. Two modes of operation are foreseen. Firstly, using a space-periodic and static magnetic field gradient. The frequency of the excitation seen by a neutron will thus vary according to its horizontal velocity. Secondly, using a homogeneous gradient oscillating in time, which should allow a more direct probe of the resonances~\cite{Pignol:2014sua}.
\begin{figure*}[!t]
 \centering
 \includegraphics[width=0.93\textwidth]{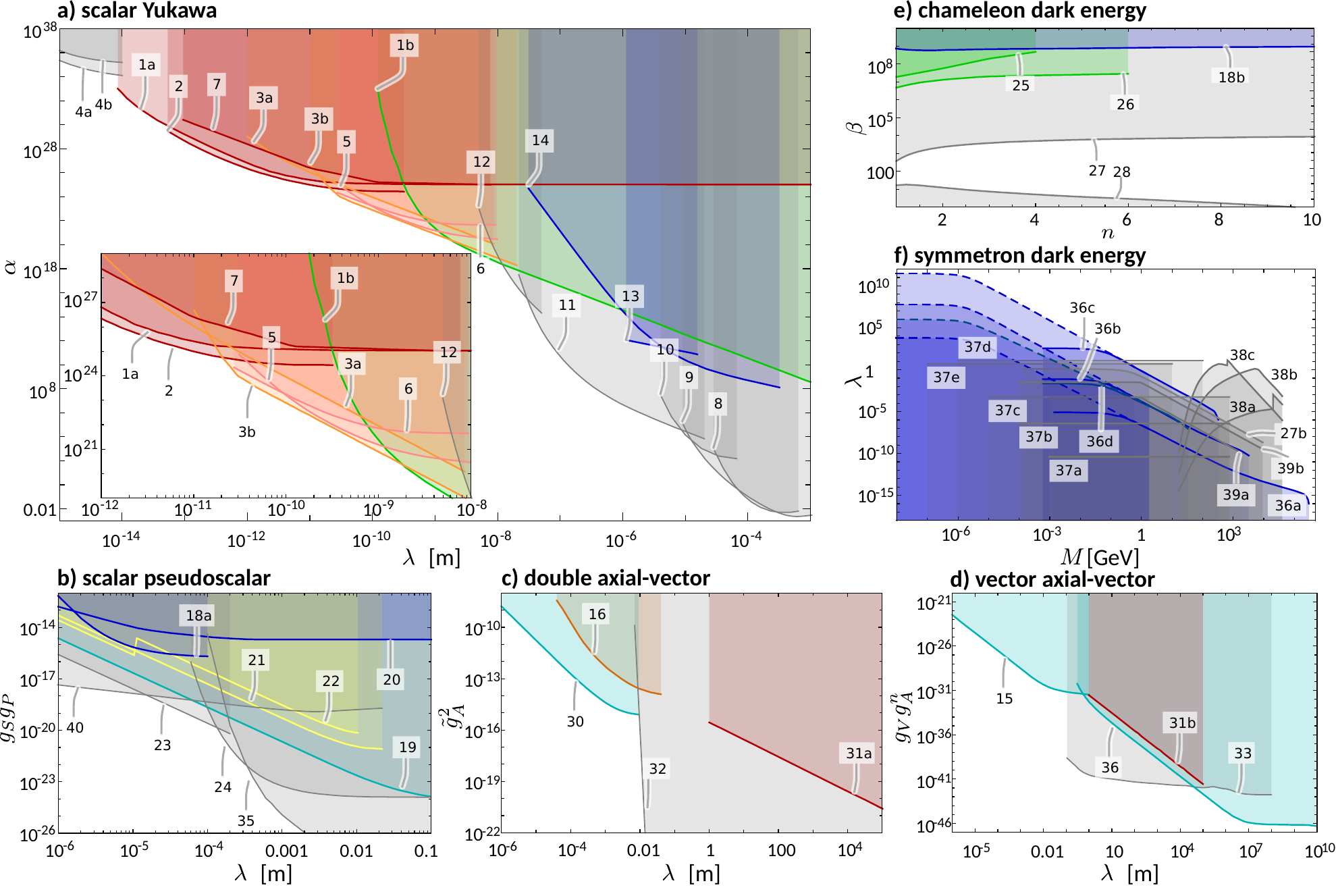}
 \caption{Compiled limits on non-Newtonian interactions. \textbf{a)} Limits on Yukawa-type modifications of Newtonian gravity. 1a and 1b: scattering and optical limits 90\% C.L.~\cite{Leeb:1992qf}, 2: neutron scattering 90\% C.L.~\cite{Pokotilovski:2006up}, 3a and 3b: Comparison of forward and backward neutron scattering, and scattering asymmetry in noble gases 95\% C.L.~\cite{Nesvizhevsky:2007by}, 5,6: Xe-neutron scattering 95\% C.L.~\cite{Kamiya:2015eva},\cite{Haddock:2017wav}, 7: silicon powder neutron scattering 67\% C.L.~\cite{Voronin:2018rmz}, 13: Gravitational quantum states 90\% C.L.~\cite{Abele:2003ga}, 14: Bouncing neutrons 95\% C.L. (repulsive)~\cite{Cronenberg:2015bol}. Non-neutron limits: 4a and 4b: nuclei charge radii and binding energies in $^{208}$Pb 67\% C.L.~\cite{Xu:2012wc}, 8: Torsion balance 95\% C.L.~\cite{Tan:2016vwu}, 9: Torsion balance 95\% C.L.~\cite{Kapner:2006si}, 10: Torsion balance 95\% C.L.~\cite{Geraci:2008hb}, 11: Micro torsion balance 95\% C.L.~\cite{Chen:2014oda}, 12: Casimir force between sinusoidally corrugated surfaces 95\% C.L.~\cite{Klimchitskaya:2017dsh};
 \textbf{b)} Scalar-pseudoscalar interactions. 18: Rabi-GRS with UCN 95\% C.L.~\cite{Jenke:2014yel}, 19: n/$^{199}$Hg comagnetometer 95\% C.L.~\cite{Afach:2014bir}, 20: UCN transmission experiment 95\% C.L.~\cite{Baessler:2006vm}, 21: Depolarization in UCN-wall interactions~67\% C.L.~\cite{Ignatovich:2009zz} (we only show their line 7), 22: Depolarization in UCN-wall interactions~95\% C.L.~\cite{Serebrov:2009ej}. Non-neutron limits: 23: $^3$He depolarization 95\% C.L.~\cite{Guigue:2015fyt}, 24: $^{129}$Xe/$^{131}$Xe NMR frequency shift 67\% C.L.~\cite{Bulatowicz:2013hf}, 35: $^3$He/$^{129}$Xe spin precession 95\% C.L.~\cite{Tullney:2013wqa}, 40: $^3$He spin relaxation 67\% C.L.~\cite{Ignatovich:2009zz}; 
\textbf{c)} `Axial vector' interactions. 16: neutron spin precession 95\% C.L.~\cite{Piegsa:2012th}, 30: neutron polarimeter 95\% C.L.~\cite{Haddock:2018}, 31a: neutron spin echo 67\% C.L.~\cite{Parnell:2020wwb}. Non-neutron limit 32: K/$^3$He comagnetometer~\cite{Vasilakis:2008yn} (curve extracted from Fig.~18 of~\cite{Safronova:2017xyt});
\textbf{d)} `Vector-axial vector' interactions. 15: neutron spin rotation in $^4$He 67\% C.L.~\cite{Yan:2012wk}, 31b: neutron spin echo 67\% C.L.~\cite{Parnell:2020wwb}. Non-neutron limits 33: Combined limits from $^3$He polarization and torsion balance with no assumptions 67\% C.L.~\cite{Adelberger:2013faa}, Note the comment in Box 1 regarding validity of $\tilde{V}_{AA}$ and $\tilde{V}_{VA}$.; 
 \textbf{e)} Chameleon interactions. 18b: Rabi-GRS 95\% C.L.~\cite{Jenke:2014yel}, 25: neutron interferometry 95\% C.L.~\cite{Lemmel:2015kwa}, 26: neutron interferometry~95\% C.L.~\cite{Li:2016tux}. Non-neutron limits 27: Atom interferometry 95\% C.L.~\cite{Jaffe:2016fsh}, 28: Torsion balance 95\% C.L.~\cite{Upadhye:2012qu} (curve taken from Fig.3b of~\cite{Jaffe:2016fsh}), 36: neutron spin rotation in $^3$He~67\% C.L.~\cite{Yan:2015};
 \textbf{f)} Symmetron interactions. 36a--d: Rabi-GRS $\mu=\{10^{-4},\,10^{-3}\,,10^{-2}\,,10^{-1}\}\,$eV, solid: Fermi-screened, dashed: micron-screened 95\% C.L.~\cite{Cronenberg:2018qxf, Jenke:2020}. Non-neutron limits 27b: Atom interferometry $\mu=0.1\,$eV 95\% C.L.~\cite{Jaffe:2016fsh}, 37a--e: Casimir experiments 95\% C.L., $\mu=\{10^{-5},\,10^{-4},\,10^{-3}\,,10^{-2},\,10^{-1}\}\,$eV~\cite{Elder:2019yyp}, 38a--c: Torsion balance $\mu=\{10^{-4},\,10^{-3}\,,10^{-2}\}\,$eV 67\% C.L.,~\cite{Upadhye:2012rc}, 39a and b: Atom interferometry $\mu=\{10^{-2}\,,10^{-1}\}\,$eV 90\% C.L.~\cite{Sabulsky:2018jma}.
 \label{fig:limits}}
\end{figure*}

The spin, being a pure quantum degree of freedom, extends the parameter space for measurements with neutrons.
As measurements are possible with and without spin-polarization, GRS is a versatile method that makes it possible to test a wide range of potentials and models including scalar-pseudoscalar axion interactions~\cite{Jenke:2014yel}, Cartan gravity~\cite{Abele:2015uua,Ivanov:2016eug}, screened dark energy chameleon and symmetron fields~\cite{Jenke:2014yel,Cronenberg:2018qxf}. GRS may also be utilized to search for a non-vanishing electric charge of the neutron~\cite{DurstbergerRennhofer:2011ap} or violations of the weak equivalence principle. The sensitivity of GRS scales with the interaction time $\tau$ of neutrons with the setup. A recent implementation of Ramsey-type GRS increases $\tau$ by a factor 4~\cite{Sedmik:2019twj}. Four orders of magnitude improvement seem possible if neutrons are stored instead of passing through the setup~\cite{Abele:2009dw}.\\
The QB has also been realized in horizontal direction, where the gravitational potential is replaced by the centrifugal potential $V(r)=rm_nv^2/2$~\cite{Nesvizhevsky:2010}. Here, $v$ is the tangential velocity of the UCNs and $r$ is the circular mirror's radius. In the same way as for the gravitational QBB, the wave functions in the radial direction are also quantized. However, the states are velocity-dependent and there is always a non-vanishing probability for tunneling through the radial barrier~\cite{Nesvizhevsky:2011}. 
\subsection{Other Searches for New Physics}
\label{sec:grav:other}
Neutron storage experiments are designed to determine the lifetime of free neutrons but can also be used to search for new physics if polarized neutrons are used. 
CP-violating interactions $\propto\bm{\sigma}\cdot\mathbf{r}$ cause a rotation of the neutron spin (represented by the Pauli spin matrix $\bm{\sigma}$) on every wall bounce. By comparing the predicted depolarization rate with the one measured in neutron bottles~\cite{Ignatovich:2009zz,Serebrov:2009ej}, limits on the respective coupling parameters $g_S g_P$ can be derived. A different way to obtain such limits in a storage chamber has been followed by the nEDM collaboration. They used Ramsey spectroscopy to measure the ratio of the precession frequencies of UCN and $^{199}$Hg atoms~\cite{Afach:2014bir} for different holding fields. Any spin-dependent interaction between the unpolarized walls and the two particle species would influence this ratio. 
The following experiments could probe vector-axial vector or double axial vector interactions but an error in the analysis has invalidated derived limits (see Box 3 and \secref{sec:grav:limits}).
Piegsa \emph{et al}~\cite{Piegsa:2012th} let a slow polarized neutron beam pass parallel to an unpolarized flat source mass. A hypothetical interaction $\propto \bm{\sigma}\cdot(\mathbf{v}\times\mathbf{r})$ would lead to an effective magnetic field normal to the neutron velocity and source plate normal vector in which the neutron spin would precess, leading to a phase shift of the observed Ramsey pattern. 
The same hypothetical interaction was tested in polarimeters~\cite{Haddock:2017wav}. Here, a neutron beam is blocked by two orthogonally oriented polarizers. Exchange of a spin 1 boson between a test mass placed between polarizers the neutron beam would induce a minute rotation of the particle spin, thereby allowing some neutrons to pass the second polarizer. From not observing such transmission, limits on the coupling parameter $\tilde{g}_A^2$ were derived. A further method to probe double axial-vector and vector axial-vector interactions is spin-echo small angle neutron scattering (SESANS)~\cite{Mezei:1972}. Here, two neutron paths are coherently separated and recombined using magnetic fields, resulting in a higher sensitivity, as compared to perfect crystal neutron interferometers relying on Bragg scattering. Since neutrons in different interferometer paths have different polarization, it was suggested that the discrepancy between interferometer data and theory was due to a hypothetical spin-dependent interaction. Parnell \emph{et al}~\cite{Parnell:2020wwb} excluded this conjecture but the analysis suffers from the assumption of an invalid potential.

Apart from hypothetical forces, tests of the Newton equivalence principle (NEP) have also been performed with neutrons. Frank \emph{et al} compared the gravitational energy $m_{n,g} g h$ of neutrons falling a height $h$ onto a rotating diffraction grating of reciprocal lattice vector $q$ with the kinetic energy $\hbar \omega+\hbar^2/(2m_{n,i})(2k_xq-q^2)$ acquired in the reflection from the grating. From knowledge of the neutron horizontal wave vector $k_x$, the change $q$ in momentum, and the rotational frequency $\omega$, limits on the ratio between inertial $m_{n,i}$ and gravitational $m_{n,g}$ mass of the neutron can be derived. 
Tests of the equivalence of $m_{n,i}$ and $m_{n,g}$ can be performed using interferometry~\cite{Colella:1975dq,Staudenmann:1980uqe}(see \secref{sec:grav:if}) or GRS. 
  Various limits from older neutron experiments and measurements at higher energies were reviewed by Dubbers and Schmidt~\cite{Dubbers:2011ns} and Abele~\cite{Abele:2008zz}.
%
%

\subsection{Experimental Limits on Non-Newtonian Interactions}
\label{sec:grav:limits}
In \figref{fig:limits}, we attempt to collect all limits obtained with neutron experiments on the various classes of Yukawa interactions~\cite{Dobrescu:2006au} with coupling parameters $\a=g_s^2\hbar c/(4\pi G_N m_n^2)$, $g_s g_p$, $\tilde{g}_A^2$, $g_Vg_A^n$, as well as for the $\beta$ parameter for chameleon interactions with $\Lambda$ fixed to the dark energy scale, and cuts $\lambda(M)$ through the three-parameter space of symmetron dark energy at different values of $\mu$. Details of the potentials can be found in Box 3. Limits from neutron experiments are given in color while other limits included for comparison are given in grey. Red and Pink indicate scattering experiments or SESANS, Orange stands for spectroscopy, Green for neutron-optical methods, turquoise for spin rotation and blue for QB and GRS.
{\red{}The most widely tested class are scalar interactions, where neutron experiments provide the tightest limits in a wide interaction range $10^{-14}\,{\rm m}<\lambda<10^{-8}\,$m as shown in \figref{fig:limits}a. Constraints for spin-dependent interactions are given in \figref{fig:limits}b. We recalculate the results from different publications to the potentials $V_s(r)$ (scalar), $V_{sp}(r)$ (scalar pseudoscalar) given in Eqns.~\eqref{eq:scalar} and \eqref{eq:scalar-pseudoscalar}. Limits on double axial-vector and vector-axial vector interactions are shown in \figref{fig:limits}c--d. For the latter interaction, we focus on the subset $g_Vg_A^n$ where the axial vector coupling is to the neutron spin~\cite{Yan:2012wk,Parnell:2020wwb} or a nucleus dominated by the neutron spin~\cite{Yan:2015,Adelberger:2013faa} and $g_V=2(g_V^e+g_V^p+g_V^n)$~\cite{Yan:2012wk}. We note that limits on $\tilde{g}_A^2$, $g_V g_A^n$ from references~\cite{Vasilakis:2008yn,Yan:2012wk,Adelberger:2013faa,Yan:2015,Haddock:2018,Parnell:2020wwb} are based on the potentials from Ref.~\cite{Dobrescu:2006au} for which some issues have been pointed out~\cite{Fadeev:2018rfl} (see Box 3).}
Strictly speaking, in order to properly relate experimental constraints to a specific theoretical model the physical details of the parts of the individual experimental setup, which act as source of the hypothetical interaction must be fully considered. Depending on the theoretical model the coupling to protons, neutrons and electron differs, which is also true for different macroscopic materials. Experimental constraints on the various non-Newtonian potentials should therefore be disentangled concerning the vertex of the interaction which may contain derivatives or not as emphasized in~\cite{Mantry:2014zsa, Mantry:2014} and reviewed in~\cite{Safronova:2017xyt}. Due to the lack of information for many experiments, we refrain from such disentanglement in the present review and rather compare limits for the same potential as given in the original publications.
Hypothetical chameleon interactions have been searched for using neutron interferometry~\cite{Lemmel:2015kwa,Li:2016tux} and GRS~\cite{Jenke:2014yel}, but neutron limits are no longer competitive. \figref{fig:limits}c shows neutron and other limits on the coupling parameter to matter $\beta$ with the potential $V(\phi)=\Lambda^4+\Lambda^{4+n}/\phi^n$ fixed by the dark energy scale $\Lambda=\Lambda_0\approx 2.4\,$meV~\cite{Brax:2004qh}. Very competitive limits on symmetron interactions have been derived from older GRS data~\cite{Cronenberg:2018qxf} (see \figref{fig:limits}d). We present cuts $\lambda(M)$ of the symmetron's three-parameter space of limits with different fixed $\mu$ as indicated.

The best current limits on the Newton equivalence principle from neutron experiments have been reported by Schmiedmayer~\cite{Schmiedmayer:1989pb}, who obtained $\gamma<1.1\pm1.7\times 10^{-4}$ by comparing NGR results for the coherent scattering length with those from scattering measurements on carbon. Frank \emph{et al.}~\cite{frankNewGravitationalExperiment2007} achieved $\gamma< 1.8\pm2.3\times10^{-3}$. These limits for the NEP are not directly comparable to the best WEP limit $\eta=2[(m_i/m_g)^\text{Pt}-(m_i/m_g)^\text{Ti}]/[(m_i/m_g)^\text{Pt}+(m_i/m_g)^\text{Ti}]\leq -1\pm27\times10^{-14}$ obtained from the MICROSCOPE space mission~\cite{Berge:2017ovy}.

\vspace{5mm}

\section{Outlook}
\label{sec:outlook}
In many fields of fundamental physics, the neutron as a massive quantum particle gives access to numerous parameters that would be hard to probe in other ways. At low energies, neutron experiments allow for accurate tests of fundamental quantum-statistical theorems and to elucidate the behavior of matter waves. 
The experiments presented here have been performed at neutron sources in different countries; we just mention the ILL (Grenoble, France), Atominstitut (Vienna, Austria), ISIS (Oxfordshire, UK), NIST (Gaithersburg, USA), and KURRI (Kyoto, Japan). These worldwide efforts are reflecting an unquenchable thirst for insights into the fundamental issues of quantum mechanics, gravity and dark energy. 
We emphasize that most of the experiments discussed in this review that test fundamental concepts of quantum mechanics have been preformed with neutron interferometry using thermal neutrons \cite{dannybook,RauchBook}, or with the quantum bouncer using ultra-cold neutrons. Tests of gravity and searches for an electric dipole moment of the neutron are predominantly performed with ultracold neutrons in gravity resonance spectroscopy. Using either higher (epi-thermal) or low (cold) energies may offer further experimental opportunities.
Another option to improve precision is to use objects of larger mass. Matter wave optics are also used in atom interferometry, which can be used to test the weak equivalence principle  with high precision \cite{Will:2014kxa}. Decoherence and interference near the classical limit has been studied with (macro-)molecules.
One particular challenging task in the future is to test possible extensions of quantum mechanics overcoming the probabilistic character of the theory. Intrinsically, quantum mechanics only allows for predictions on the entire ensemble and does not give information about individual "events". Besides the classical Einstein-Podolsky-Rosen approach \cite{EPR35}, possible extensions include the time-symmetric interpretation of quantum mechanics \cite{Aharonov64,Aharonov95} or Bohmian mechanics \cite{Bohm52}. However, to date no experiment that can distinguish between quantum mechanics and such alternative theories exists.

The neutron is also the smallest practically available massive electrically neutral particle that allows for accurate tests of Newtonian gravity and possible modifications thereof. Neutron scattering still gives the tightest limits on scalar Yukawa interactions in the range \SI{10}{\femto\metre}--\SI{10}{\nano\metre}, but both theory and experiment are accurate at the $10^{-4}$ level in measured scattering lengths. Any improvement requires advancements on the experimental \emph{and} theoretical side. However, neutron-nucleon interactions remain hard to model and on the experimental side isotopic composition of scattering target, perfect control of various environmental parameters~\cite{Voronin:2018rmz}, and background~\cite{Haddock:2017wav,Pokotilovski:2006up} are limitations. The combination of data sets may statistically improve limits~\cite{Nesvizhevsky:2007by}. Snow \emph{et al}~\cite{Snow:2019mbb} speculate that accurate definition of the falling height in  Neutron gravity refractometry could improve scattering length measurements by 1--2 orders, which would improve limits on Yukawa interactions proportionally, as the theory is more under control here. In any case, the interaction range mentioned can at present only be covered by neutron experiments. For scalar-pseudoscalar interactions, neutron co-magnetometry may yield some improvements~\cite{Afach:2014bir} below $\SI{100}{\micro\metre}$, while at larger interaction ranges, atom-based methods~\cite{Tullney:2013wqa,Bulatowicz:2013hf} are clearly more sensitive.
{\red{}Prospects for vector-axial vector and double axial vector interactions have to be evaluated after careful re-analysis of the potentials involved~\cite{Fadeev:2018rfl}. However, as neutrons already give the tightest limits for $\tilde{g}_{AA}$ at $\lambda\lesssim\SI{1}{\centi\metre}$, it may be interesting to look at measurements where both the neutron and the target (for example an extended mass distribution) are spin-polarized, as this would make it possible to probe the full $V_{AA}$ and $V_{AV}$~\cite{Fadeev:2018rfl} that are not suppressed by $v/c$.}
With respect to dark energy models, gravity resonance spectroscopy 
provides best limits for symmetron interactions~\cite{Cronenberg:2018qxf,Jenke:2020} on a large parameter space and bears substantial potential for improvement. On the other hand, both neutron interferometry and GRS are not competitive at the moment for chameleon interactions but the situation changes if neutrons are stored~\cite{Abele:2009dw}.
GRS has been proposed as a test of Einstein-Cartan gravity~\cite{Abele:2015uua}, beyond-Riemann gravity~\cite{Ivanov:2021bvk} entropic gravity~\cite{Schimmoller:2020kvg}, the Newton equivalence principle, and the weak equivalence principle. The latter would represent the first test of the weak equivalence principle with hadrons and would, independently of the achievable precision, represent a class of its own. Ref.~\cite{Voronin:2011} proposed testing the Newton equivalence principle using Bragg-diffraction in large crystals at large angle anomalous absorption has hampered the extraction of limits so far~\cite{Voronin:2017fut}. As scattering experiments are sensitive to a wide range of energies, the possibility of detecting extra dimensions with neutrons was discussed~\cite{Frank:2003ms}.

\vspace{3mm}

In conclusion, new, unexpected aspects of particle-wave duality may emerge from neutron interferometry. Furthermore, prospects for the exclusion or possible detection of hypothetical interactions are bright in next-generation neutron gravity experiments.


\vspace{5mm}

\subsection*{Acknowledgements}
The authors thank our co-workers and collaborators for their long-term efforts and support, especially we want to thank the Institut Laue-Langevin (ILL), in Grenoble France, for ongoing support and hospitality. This work was supported by the Austrian science fund (FWF) Project Nos. P 30677 and P 27666. Y. H. is partly supported by KAKENHI.


\end{document}